\documentclass{article}

\usepackage[final,nonatbib]{Seed-ASR}

\usepackage{graphicx}
\usepackage{pdfpages}
\usepackage{subfigure}

\usepackage[utf8]{inputenc} 
\usepackage[T1]{fontenc}    
\usepackage{hyperref}       
\usepackage{url}            
\usepackage{booktabs}       
\usepackage{amsfonts}       
\usepackage{nicefrac}       
\usepackage{microtype}      
\usepackage{xcolor}         
\usepackage{makecell}
\usepackage{multirow}
\usepackage{setspace}

\usepackage{soul} 
\usepackage{xcolor} 
\sethlcolor{gray}

\usepackage{enumitem}
\usepackage{bm}
\usepackage{amsmath}
\definecolor{lightgray}{RGB}{230,230,230}

\title{Seed-ASR: Understanding Diverse Speech and Contexts with LLM-based Speech Recognition}

\author{%
  Seed Team, ByteDance\thanks{Please cite this work as "Seed-ASR (2024)". The list of authors can be found at the end of the document.} \\
}

\begin{document}

\maketitle

\begin{abstract}
\setcounter{footnote}{0}
Modern automatic speech recognition (ASR) model is required to accurately transcribe diverse speech signals (from different domains, languages, accents, etc) given the specific contextual information in various application scenarios. Classic end-to-end models fused with extra language models perform well, but mainly in data matching scenarios and are gradually approaching a bottleneck. In this work, we introduce Seed-ASR\footnote{Seed-ASR capabilities have been applied in a variety of ByteDance products, and have provided technical commercialization services in China with Doubao speech recognition models.}, a large language model (LLM) based speech recognition model. Seed-ASR is developed based on the framework of audio conditioned LLM (AcLLM), leveraging the capabilities of LLMs by inputting continuous speech representations together with contextual information into the LLM. Through stage-wise large-scale training and the elicitation of context-aware capabilities in LLM, Seed-ASR demonstrates significant improvement over end-to-end models on comprehensive evaluation sets, including multiple domains, accents/dialects and languages. Additionally, Seed-ASR can be further deployed to support specific needs in various scenarios without requiring extra language models. Compared to recently released large ASR models, Seed-ASR achieves 10\%-40\% reduction in word (or character, for Chinese) error rates on Chinese and English public test sets, further demonstrating its powerful performance.
\end{abstract}

\begin{spacing}{1.0}
\section{Introduction}
\begin{figure}[h] 
\centering 
\includegraphics[width=1\textwidth, trim=4cm 1cm 4cm 0cm, clip]{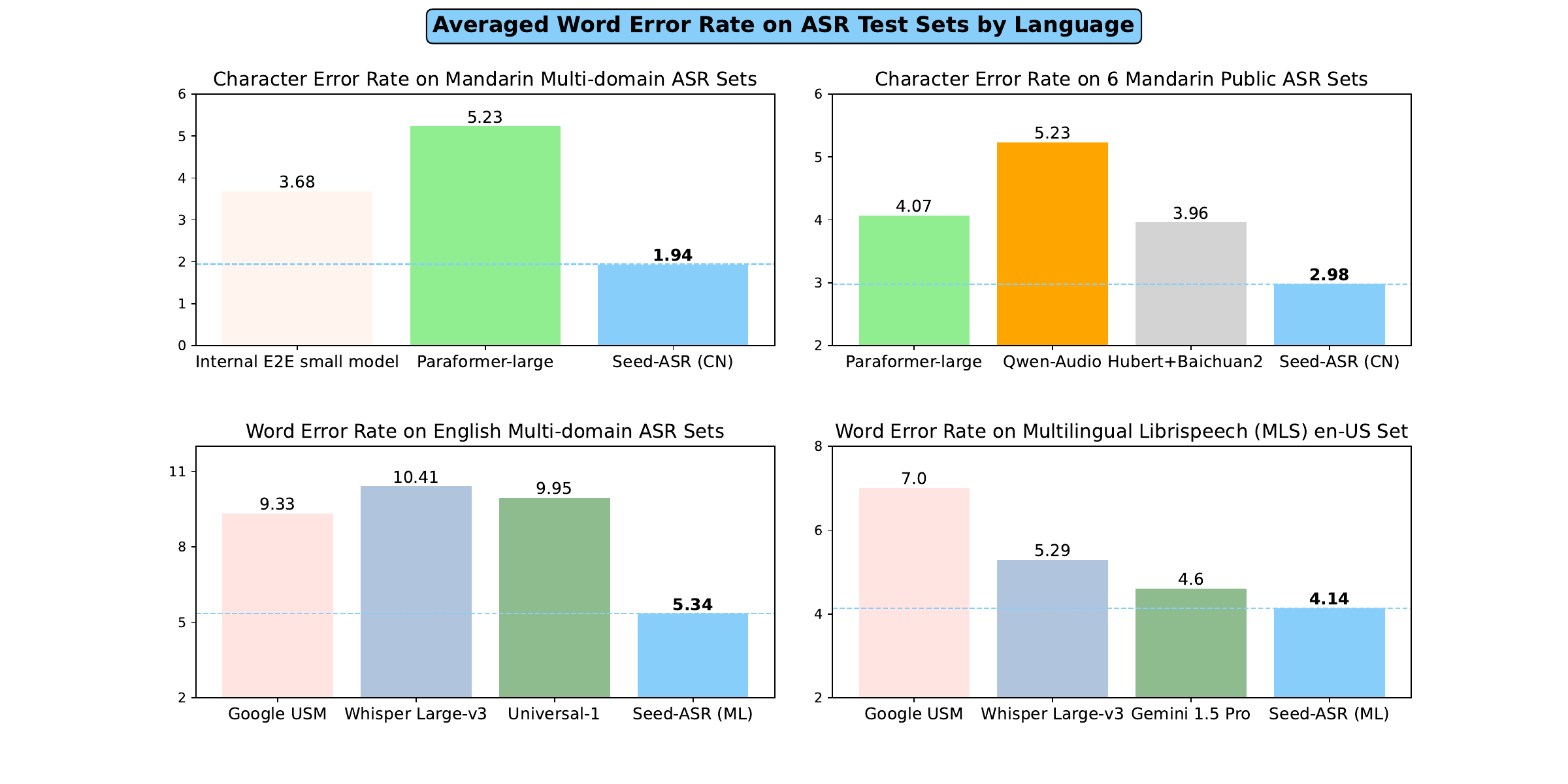}
\caption{The comparison of ASR performance between Seed-ASR and other strong released models on our internal multi-domain evaluation sets and public sets, covering both Mandarin and English. The MLS en-US result of Whisper Large-v3 is obtained by locally decoding the MLS en-US test set because there is no reported WER on published papers or technical reports.}
\label{acc} 
\end{figure}

We present Seed-ASR, an LLM-based large-scale ASR model. Aiming to become a "smarter" speech recognition model, Seed-ASR is developed under the framework of audio conditioned LLM (AcLLM), leveraging the capability of LLMs by inputting continuous speech representation together with instruction and contextual information into the LLM. Seed-ASR has five major features: 

\begin{enumerate}
\item \textbf{High Recognition Accuracy}: By training on over 20 million hours of speech data and nearly 900 thousand hours of paired ASR data, our Chinese multi-dialect model, Seed-ASR (CN), and our multilingual model, Seed-ASR (ML), achieve impressive results on public datasets and our in-house comprehensive evaluation sets (shown in Figure \ref{acc});
\item \textbf{Large Model Capacity}: Seed-ASR employs an audio encoder with nearly 2 billion parameters and a Mixture of Experts (MoE) LLM with tens of billions of parameters for modeling. The experiments of scaling law on ASR tasks underpin our decision to choose large models;
\item \textbf{Multiple language Support}: While maintaining high accuracy, Seed-ASR (CN) supports transcribing Mandarin and 13 Chinese dialects with a single model. Additionally, Seed-ASR (ML) recognizes speech of English and 7 other languages, and is being extended to support more than 40 languages;
\item \textbf{Context-aware Ability}: Seed-ASR utilizes a range of contextual information, including historical dialogues, video editing history, and meeting participation details, in a unified model to capture essential indicators related to speech content. This integration substantially boosts keyword recall in ASR evaluation sets across various scenarios.
\item \textbf{Stage-wise Training Recipe}: The development of Seed-ASR goes through a simple and effective training recipe: self-supervised learning (SSL) of audio encoder $\rightarrow$ supervised fine-tuning (SFT) $\rightarrow$ context SFT $\rightarrow$ reinforcement learning (RL). Each stage has a distinct role, ensuring stage-by-stage performance improvement of Seed-ASR.
\end{enumerate}

Different from existing LLM-based ASR models \cite{tang2023salmonn,chen2023x,rubenstein2023audiopalm,huang2024audiogpt,wu2023decoder,li2023prompting,chu2023qwen}, Seed-ASR aims for extensive improvements in ASR performance over the state-of-the-art in ASR technology across multiple languages including Chinese and English, tailored for a broad array of application scenarios featuring varied speech types and contexts. To achieve this, we build a series of high-quality evaluation sets that include a wide range of speech inputs, such as different domains, accents/dialects, languages and speech duration. These sets also cover evaluation of the customization capability of an ASR system under different application scenarios (e.g. the keyword accuracy and consistency in conversations). In designing Seed-ASR, we chose the path of large-scale training, leveraging both substantial model capacity and extensive training data to enhance generalization ability. We also elaborate the customization capability by training the model to account for contexts provided to the AcLLM framework, forming a unified and compact model structure for different scenarios. On our multi-dimensional evaluation sets, Seed-ASR demonstrates more comprehensive and powerful model capability compared to the classic end-to-end models. The performance advantage of Seed-ASR is further evidenced in public test sets and our subjective understanding evaluations. In the following sections, we will introduce our motivation, methods, models and evaluation in detail.

\section{Motivation}
Since the rise of neural networks (NNs) and deep learning in 2010s, the modeling of automatic speech recognition (ASR) has experienced an upgrade from the hybrid framework \cite{hinton2012deep,graves2006connectionist,miao2015eesen} that only relies on NN-based acoustic models to the end-to-end (E2E) framework \cite{graves2012sequence,bahdanau2016end,chan2016listen,watanabe2017hybrid,dong2020cif} in which the entire NN models are trained to output transcriptions directly. Although significant progress has been made in recognition accuracy as measured by word error rate (WER), current end-to-end ASR models are still not "smart" enough, which is limited by the model capacity and from-scratch training manner. Specifically, it cannot efficiently utilize rich common sense knowledge and conduct contextual reasoning during the recognition process, thus inevitably relying on the complicated fusion strategy with extra language models. With the rapid development of LLM technology \cite{brown2020language,openai2022chatgpt,openai2023gpt4,chowdhery2023palm,anil2023palm,touvron2023llama,touvron2023llama2}, the potential of AI continues to grow. Automatic speech recognition (ASR), as a classic task in AI, also stands at the brink of advancements in its model framework.

The upgrade of the ASR model could get inspirations from the technical advancements of LLM, which can be attributed to three main aspects: 

\begin{itemize}
\item Unified model framework. LLM employs a decoder-only framework based on the next-token-prediction. It sequences the input and output text, relying on the self-attention mechanism to establish dependencies between tokens in sequences, thereby unifying text understanding and text generation;
\item The power of scaling law. Large-scale model parameters provide crucial capacity for LLM to learn knowledge from diverse data sources. For example, from GPT-2 \cite{radford2019language} to GPT-3 \cite{brown2020language}, the number of parameters increases from 1.5 billion to 175 billion, enabling GPT-3 to exhibit better generalization and emergent abilities;
\item Comprehensive training pipeline, ChatGPT goes through three stages: pre-training, supervised fine-tuning (SFT) and reinforcement learning with human feedback (RLHF). In the stage of pre-training, LLM is trained on a large amount of text data, which makes it store extensive knowledge. In the stage of SFT, LLM is further fine-tuned on higher-quality, task-oriented data, enhancing its ability to reason with context and understand task instructions. Finally, in the RLHF stage, the training objective shifts to align the LLM's behavior with human preferences with the help of reinforcement learning; 
\end{itemize}

Since the task of ASR is to convert speech to text, its text generation process is consistent with that of LLMs. The extensive text knowledge and contextual reasoning capabilities stored in LLMs make them potential components for providing semantic guidance to ASR. The remaining core challenge is how to enable LLMs to better "understand" speech, a modality that is different from text.

\section{Methods}
\begin{figure}[] 
\centering 
\includegraphics[width=\textwidth]{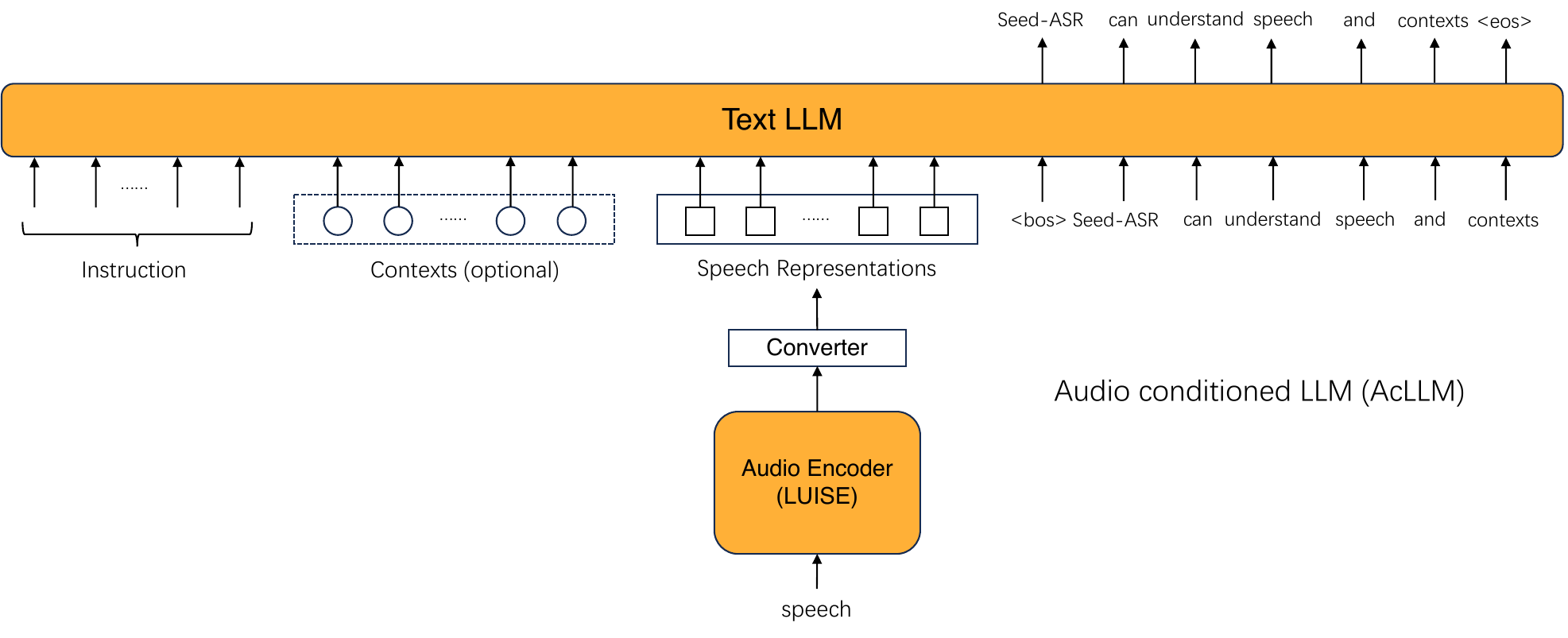}
\caption{The model framework used in Seed-ASR. When contexts are provided, the instruction is "There are relevant contexts, transcribe the speech into text:". Otherwise, the instruction is "Transcribe the speech into text:". }
\label{AcLLM_framework} 
\end{figure}

\subsection{Framework and Training Recipe}
Based on the aforementioned motivation, we propose Seed-ASR, a large-scale speech recognition model built on the framework of audio conditioned LLM (AcLLM). By inputting encoded continuous speech representations together with a task instruction and relevant contexts into a pretrained LLM, Seed-ASR can leverage the rich text knowledge and the reasoning ability of the LLM to generate the corresponding text transcription of speech. The overall framework is shown in Figure \ref{AcLLM_framework}

\begin{figure}[] 
\centering 
\includegraphics[width=0.8\textwidth]{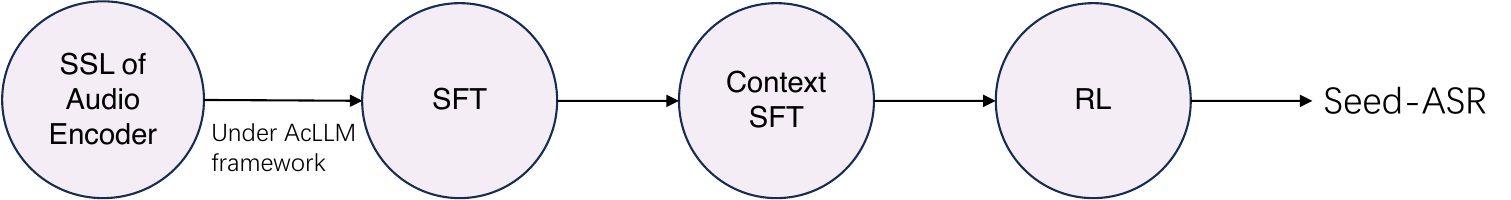}
\caption{The stage-wise training recipe for the development of Seed-ASR. SSL represents self-supervised learning, SFT represents supervised fine-tuning, RL represents reinforcement learning.}
\label{train_stage} 
\end{figure}

Audio is a different modality from text. To enable LLMs better understand diverse speech inputs, we adopt the concept of large-scale pretraining in LLMs. Specifically, we construct an audio encoder with nearly 2 billion parameters and conduct self-supervised learning (SSL) on tens of millions of hours of data. 
The pre-trained audio encoder gains strong speech representation ability, which facilitates rapid convergence during supervised fine-tuning (SFT). After the large-scale SSL stage, we implement a simple and effective stage-wise training recipe within the framework of AcLLM (shown in Figure \ref{train_stage}). In the stage of SFT, we establish the mapping relationship between speech and text by training on a large amount of speech-text pairs. In the stage of context SFT, we use a relatively small amount of context-speech-text triples to elicit the LLM's ability to capture speech-relevant clues from context. These triple data can be customized according to specific scenarios. In the stage of reinforcement learning, we apply the training criteria of MWER \cite{prabhavalkar2018minimum} and some improvements to further strengthen the ability of our model. In the following subsections, we will introduce these methods in more detail.

\subsection{SSL of Audio Encoder}
Large-scale SSL enables audio encoders to capture rich information from speech. Inspired by the BERT-based speech SSL framework \cite{hsu2021hubert,baevski2020wav2vec,chen2022wavlm,chiu2022self}, we developed our audio encoder, a conformer-based \cite{gulati2020conformer} model that captures both global and local structures stored in audio signals. In this work, we primarily focus on speech signal. Since it is trained on large-scale unsupervised data, we term the trained audio encoder as 
LUISE, which represents \textbf{L}arge-scale \textbf{U}nsupervised \textbf{I}terative \textbf{S}peech \textbf{E}ncoder.

\begin{figure}[h] 
\centering 
\includegraphics[width=0.45\textwidth]{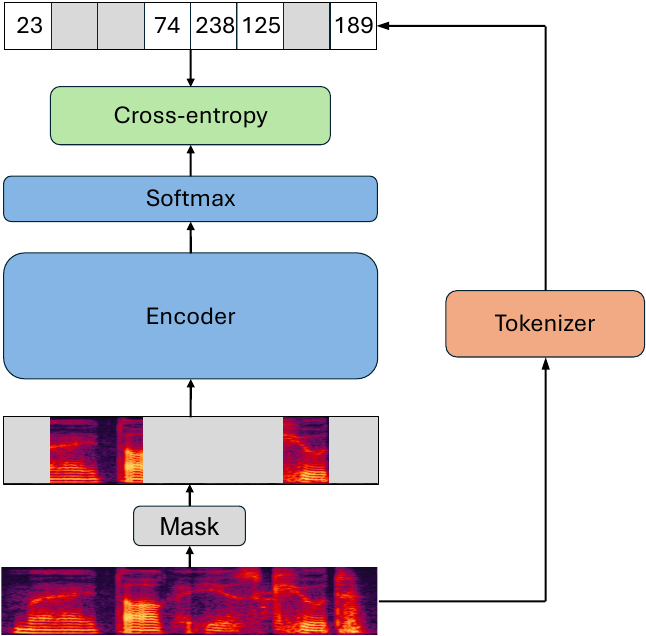}
\caption{The training procedure of our audio encoder LUISE.}
\label{LUISE} 
\end{figure}

Adhering to the concept of BERT \cite{devlin2018bert}, LUISE adopts a learning paradigm of masked language prediction. The training procedure is illustrated in Figure \ref{LUISE}. Specifically, the sequence of mel-filterbank feature extracted from the waveform is first input to the tokenizer module to obtain discrete labels for each frame. Then, the training of LUISE is conducted using the cross-entropy criterion, with the loss function calculated only for the masked frames. After training, the softmax layer is removed, and the encoder part of LUISE is used for subsequent supervised fine-tuning.

\begin{figure}[] 
\centering 
\includegraphics[width=0.6\textwidth]{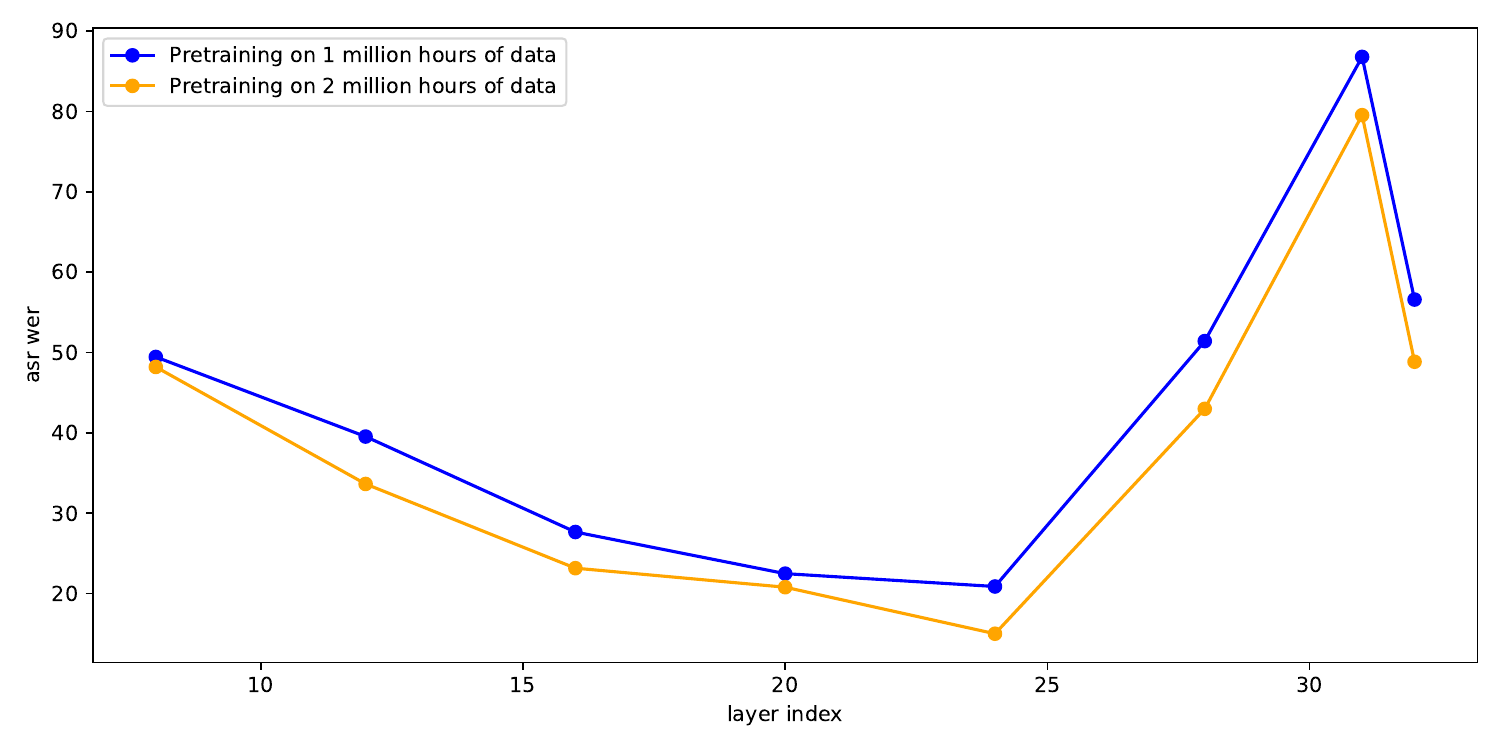}
\caption{The probing experiment of the layer with the best semantic representations in LUISE. The result of word error rate is obtained by conducting greedy search with a CTC model. }
\label{layer_probing} 
\end{figure}

We utilize an iterative fixed tokenizer method to obtain the corresponding discrete labels for each frame. In the first iteration, we apply a random-projection layer \cite{chiu2022self} to project speech feature to a randomly initialized codebook, and map them to discrete labels through finding the nearest vector in the codebook. In the second iteration, we perform K-means clustering on the representations of an intermediate layer of the previously trained encoder to obtain a new codebook. The discrete labels are then obtained by finding the closest vector in the new codebook to the representation from the same intermediate layer. During the selection of the intermediate layer, we freeze the parameters of encoder trained in the first iteration, and add a mapping layer and connectionist temporal classification (CTC) \cite{graves2006connectionist} loss to each intermediate layer for supervised fine-tuning. Figure \ref{layer_probing} shows the word error rate (WER) obtained from supervised fine-tuning on the representation of each intermediate layer. For LUISE with 2 billion parameters, the output at the 25th layer (out of 32 layers) demonstrates the best semantic representation and is used for the generation of discrete labels in subsequent iterations.

\subsection{SFT}
After the training on large-scale speech-only data, LUISE has developed strong speech representation capabilities. It outputs continuous representation containing rich speech and semantic information at a frame rate of 40ms. In order for AcLLM to better understand the corresponding text content in speech, we need to map the semantic information from the encoded representation into the semantic space of the LLM. To achieve this, we use the following two methods:

\begin{itemize}[left=2pt]
\item In the model structure, we introduce a converter module to connect our audio encoder (LUISE) with the LLM (as shown in Figure \ref{AcLLM_framework}). The converter includes a downsampling module and a linear projection layer. We find that different downsampling methods work equally well, so we utilize the most concise method: frame splicing. Specifically, we input 4 consecutive frames of speech representation to the linear layer after splicing them in the feature dimension. Consequently, the frame rate of the speech representations in Figure \ref{AcLLM_framework} inputted to the LLM is 160ms;
\item In terms of the training method, we adopt the strategy of "learnable audio encoder + learnable converter + fixed LLM", which maximizes the retention of the LLM's rich semantic knowledge and reasoning abilities by keeping its parameters unchanged. The learnable audio encoder and converter parameters ensure that the semantic information contained in the speech representation is aligned to the semantic space of the LLM. During the training process, the cross-entropy loss function is used, with only the token positions that generate the transcribed text participating in the cross-entropy calculation;
\end{itemize}

\subsection{Context SFT}
\begin{figure}[] 
\centering 
\includegraphics[width=\textwidth]{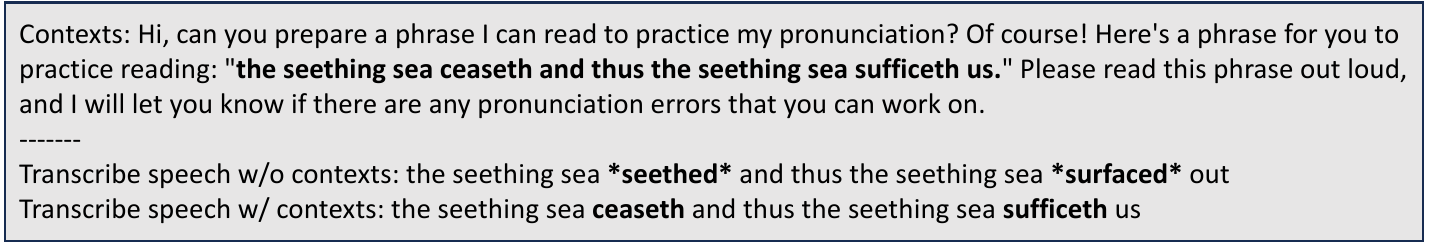}
\caption{An example of transcribing speech with or without contexts.}
\label{context_example} 
\end{figure}

After training on large-scale speech-text pair data, our SFT model achieves strong performance on test sets covering multiple domains. However, the training manner of the SFT model determines that it lacks the ability to recognize ambiguous speech content given contextual information (contexts). These issues are more pronounced in scenarios involving accents (with speech ambiguity), and homonyms or rare words (with semantic ambiguity). Therefore, we introduce context-aware training and the method of joint beam search to enhance the model's ability to utilize context effectively (an example is present in Figure \ref{context_example}). 

\begin{itemize}[left=2pt]
\item Context-aware training: First, we use our internal LLM to generate contexts related to the transcription of speech. It performs better than using the history transcription in long-form speech as the contexts \cite{radford2023robust} in our experiments. Using the generated natural language contexts could also provide more complete semantics than sampled words in \cite{chen2024salm}, thus enabling the learning of reasoning in addition to copying the relevant transcription content from contexts. Then, we build a dataset of <context, speech, text> triples, which are mixed with a certain proportion of general ASR data (speech-text pair data) for context-aware training. As shown in Figure \ref{AcLLM_framework}, during context-aware training, we input the contexts and speech representations into the LLM. The goal of this training is to enhance the model's ability to capture speech content-related clues from the contexts.

\item Joint beam search: We find that directly using the native beam search suffers from serious hallucination problem.
To address this, we propose a decoding strategy of joint beam search to alleviate this problem. Specifically, we use joint beam search to find the optimal score $P_{\text{joint}} (\bm{y} |\bm{x},\bm{c})$, where $\bm{y}$ represents the predicted hypothesis, $\bm{x}$ is the speech information, and $\bm{c} $ is the given contextual information. The hyper-parameter $\alpha$ is used to balance the importance of speech information and contextual information during the decoding:

\begin{equation}
P_{\text{joint}}(\bm{y}|\bm{x}, \bm{c}) = \frac{\alpha}{1 + \alpha} * P(\bm{y}|\bm{x}, \bm{c}) + \frac{1}{1 + \alpha} * P(\bm{y}|\bm{x})
\end{equation}

Simultaneously, we introduce a pruning strategy that first uses context-independent score $P(\bm{y} |\bm{x})$ to filter out acoustically implausible candidate tokens, and then applies joint beam search to the remaining candidate tokens. The pruning strategy plays an important role in alleviating hallucination.
\end{itemize}

\subsection{RL}
Since the training in the SFT and Context SFT stages is based on the cross-entropy objective function, there is a mismatch with the evaluation metrics used during inference (e.g. WER). With the successful development of reinforcement learning (RL), it can learn relatively optimal decision-making strategies in sequence modeling tasks. Therefore, we introduce the RL stage by constructing a reward function based on ASR metrics.

Word error rate (WER) is often considered a core metric for evaluating the performance of ASR models, but certain parts of content (e.g. keyword) in a sentence plays a more crucial role in the understanding of the whole sentence. Therefore, we also introduce the metric of weighted WER (WWER) as an additional reward function, emphasizing the importance of keyword errors.  Specifically, we apply minimum word error rate (MWER) \cite{prabhavalkar2018minimum} as another training objective interpolated with the cross-entropy objective $\mathcal{L}_{\text{CE}}$ in our RL stage:

\begin{equation}
 \mathcal{L}^{\text{N-best}}_{\text{mwer}}(\bm{x}, \bm{y^*})=\frac{1}{N}\sum_{\bm{y_i} \in \text{N-best(x, N)}} \hat{P}(\bm{y_i}|\bm{x})(\mathcal{W}(\bm{y_i},\bm{y^*})-\bar{W}) + \lambda\mathcal{L}_{\text{CE}}
\end{equation}

where $\mathcal{W}(\bm{y^*}, \bm{y_i})$ represents the WER value or WWER value (where the weight of the keyword error is increased) between the ground-truth ($y^*$) and each hypothesis $\bm{y_i}$ in $\text{N-best}(x, N)$. $\bar{W}$ represents the average WER or WWER of N-best hypotheses. $\lambda$ is the interpolation coefficient. $\hat{P}(\bm{y_i}|\bm{x})$ represents the normalized likelihood probability of hypotheses, which is calculated as follows:

\begin{equation}
 \hat{P}(\bm{y_i}|\bm{x})=\frac{P(\bm{y_i}|\bm{x})}{\sum_{\bm{y_i} \in \text{N-best(x, N)}} P(\bm{y_i}|\bm{x})}
\end{equation}

To improve the training efficiency of RL, we deploy a remote service to generate hypotheses and simultaneously calculate the MWER loss while updating the model parameters on the current server. During the RL training process: 1) we initialize the model parameters with the context SFT model trained from the previous stage; 2) we utilize high-quality data for reinforcement learning training, with a data scale of thousands of hours. 3) to preserve the context-aware capability of the initialization model, our training data also includes a certain proportion of <context, speech, text> triples. After completing the RL training, we obtain our Seed-ASR model.

\begin{table}[h]
\caption{Ablation studies in the stage of RL. Weighted WER as the reward function shows better performance than WER on all three evaluation sets (details of these sets are introduced in Section \ref{sec:seed_asr_cn}). The training data of <contexts, speech, text> triples in RL stage ensure the context-awareness ability does not drop. Seed-ASR utilizes the strategy in the last row. The metric of WER or weighted WER calculates the character error for Chinese, Japanese and Korean, and word error for English and other languages.}
\label{rl_ablation_study}
\centering
\begin{tabular}{l|c|c|c}
\toprule
Model     & Multidomain WER $\downarrow$ & Hardcase (F1\%) $\uparrow$ & \makecell[c]{Context Strict \\ (Recall\%) $\uparrow$} \\
\midrule
Model after Context SFT  &  2.02  &  93.39  &  80.63  \\
+ RL w/ WER reward   &  1.98  &  93.39  &  75.34  \\
+ RL w/ Weighted WER reward  &  1.94  &  \textbf{93.78}  &  78.01  \\
\qquad + train w/ context &  \textbf{1.94}  &  93.72  &  \textbf{80.63} \\
\bottomrule
\end{tabular}
\end{table}

\subsection{Observations}
In the process of improving the performance of Seed-ASR, we have also obtained some observations:

\subsubsection{Scaling Law}
\label{scaling_law}
In the realm of LLM, it is observed that larger models can continuously reduce the loss value by training on more data \cite{kaplan2020scaling,hoffmann2022training}. To the best of our knowledge, there is no relevant research on scaling laws for audio encoders under LLM-based framework. During the SSL stage, we conduct experiments to explore the performance of LUISE with different model sizes. Specifically, we select five groups of model sizes: 75M, 0.2B, 0.6B, 2B, and 5B. The training data comprises of 7.7 million hours of unsupervised speech-only data covering multiple domains, ensuring the full utilization of the model capacity. Different-sized models maintain consistency in most training configurations, except that as we increase the model size, we proportionally expand the width and depth of the model, appropriately increase the batch size and weight decay, and reduce the learning rate. 

\begin{figure*}[htbp]
\centering
\subfigure[]
{
\begin{minipage}[t]{0.33\linewidth}
\centering
\includegraphics[width=2.0in]{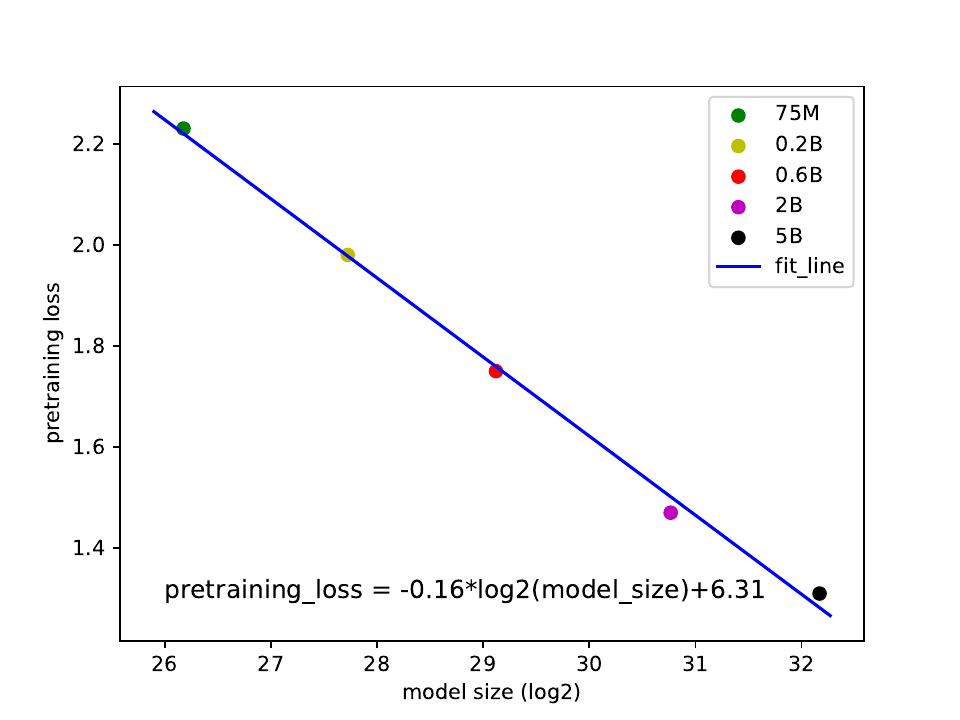}
\label{scaling_law1}
\end{minipage}%
}%
\subfigure[]{
\begin{minipage}[t]{0.33\linewidth}
\centering
\includegraphics[width=2.0in]{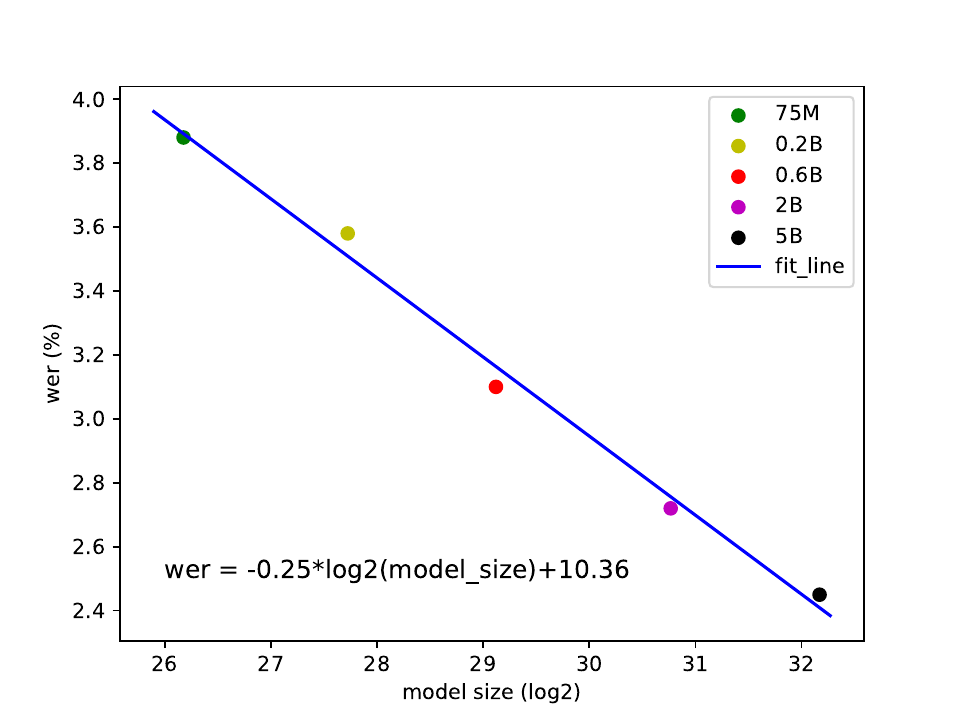}
\label{scaling_law2}
\end{minipage}%
}%
\subfigure[]{
\begin{minipage}[t]{0.33\linewidth}
\centering
\includegraphics[width=2.0in]{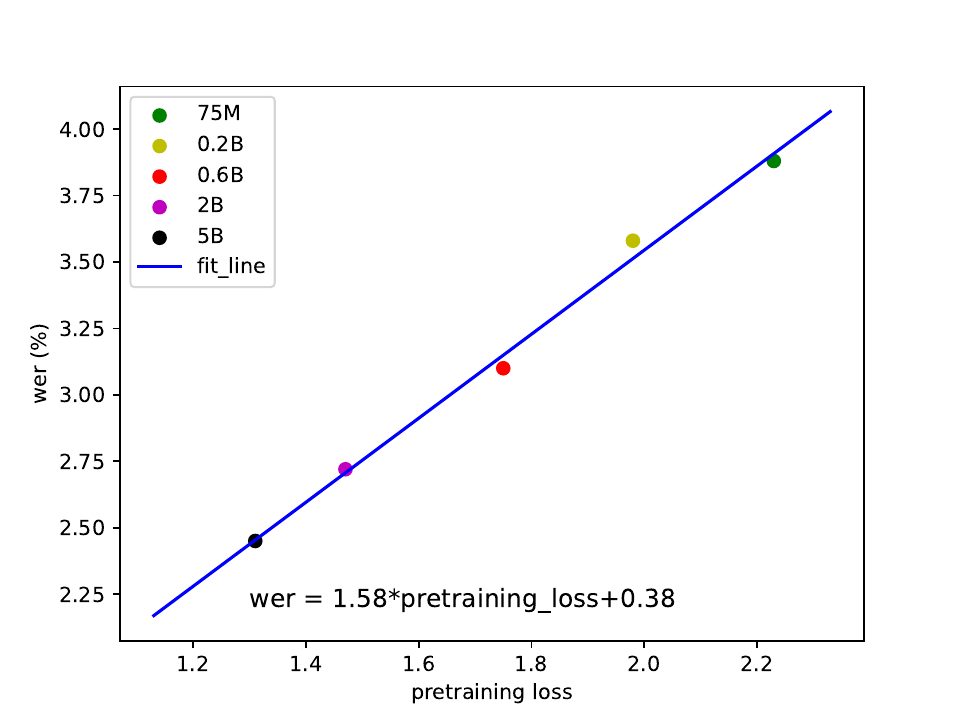}
\label{scaling_law3}
\end{minipage}
}%
\caption{\label{scaling}
(a) depicts the correlation between the pretraining loss of our audio encoder (LUISE) and base-2 logarithm of the model parameter size. (b) depicts the correlation between the greedy WER after the SFT and base-2 logarithm of the model parameter size. (c) depicts the correlation between the greedy WER after SFT and the pretraining loss of LUISE.}
\end{figure*}

We first focus on the correlation between the cross-entropy pretraining loss value on the validation set and the model size. As shown in Figure \ref{scaling_law1}, we observed a nearly linear correlation between the two. Additionally, we compared the performance after training on a small-scale SFT data based on the trained LUISE. Greedy search was used for inference. As shown in Figure \ref{scaling_law2}, the WER metric on the multidomain evaluation set also exhibits a nearly linear correlation with the model size of LUISE. Furthermore, this reveals a positive correlation between the WER metric on the test set after SFT and the loss function value in the SSL stage in Figure \ref{scaling_law3}. These findings on scaling law provide guidance for our encoder selection (taking into account the balance of performance and efficiency) and subsequent optimization.

\subsubsection{Long-form Ability}
Our Seed-ASR is modeled under the framework of AcLLM, which naturally leverages the semantic knowledge and long-context modeling capabilities of LLM. Therefore, we also explore the option of directly inputting the entire long-form speech into LLM for recognition. This approach effectively avoids two problems associated with segmenting long-form speech for multiple independent inferences: 1) The segmentation process may result in the loss of information at the boundaries, decreasing recognition accuracy; 2) The segmentation process disrupts the strong global context information in long-form speech, affecting both the accuracy and consistency of recognition. 

\begin{table}[h]
\caption{The comparison of performance on long-form video test sets. }
\label{long-form_ablation_study}
\centering
\resizebox{\textwidth}{!}{
\begin{tabular}{l|c|c|c|c|c|c}
\toprule
Model     & Avg WER & video\_1 & video\_2 & video\_3 & video\_4 & video\_5  \\
\midrule
Transducer-based E2E Model  &  3.92  &  2.83  &  3.80  &  3.80  &  4.22  &  4.66   \\
Paraformer-large  &  5.97  &  5.78  &  5.36  &  5.80  &  6.87  &  5.96  \\
Our Model after short-form SFT   &  2.28  &  1.48  &  1.99  &  2.31  &  2.64  &  2.73  \\
\qquad + long-form SFT    &  \textbf{2.08}  &  \textbf{1.44}  &  \textbf{1.96}  &  \textbf{1.95}  &  \textbf{2.56}  &  \textbf{2.31}  \\
\bottomrule
\end{tabular}
}
\end{table}

Specifically, we build a series of long-form video test sets comprising 5 datasets from different sources. During training, the entire long-form data is inputted into the model without any segmentation processing. The duration distribution of the test set is comparable to that of the training set. As shown in Table \ref{long-form_ablation_study}, using long-form data for both training and testing results in relative WER reduction of nearly 8.8\% compared to short-form training, which employs a domain-adaptive VAD to segment long-form speech into several parts for training and testing. The maximum duration of the long-form video test sets is 5 minutes, with scheduler for significant length extension.

\section{Model and Evaluation}
At present, we focus on the comprehensive improvement of Chinese and multilingual (without Chinese) speech recognition performance in diverse scenarios. Therefore, we present two Seed-ASR models with the same model structure and training recipe: the Chinese multi-dialect model, termed Seed-ASR (CN), and the multilingual model, termed Seed-ASR (ML). While we also have models that support both Chinese and multilingual languages, this report will specifically detail the two Seed-ASR models that focus on Chinese and multilingual (excluding Chinese), respectively. 

Seed-ASR (CN) not only transcribes Mandarin and 13 Chinese dialects with a single model but also demonstrates significant performance improvements over other released large models on the multi-dimensional evaluation sets, including multi-domain, multi-dialect, multi-accent and public set. Additionally, the training in the context SFT stage endows Seed-ASR (CN) with effective context-aware ability as demonstrated on dialogue context evaluation sets. Similarly, Seed-ASR (ML) achieves competitive results compared to other released models on 8 multilingual public sets (including English) and multi-domain evaluation sets, and it is being extended to more than 40 languages. The metric of word error rate (WER) is used as the main objective metric in the following part. Unless otherwise specified, the metric of WER calculates the character error for Chinese, Japanese, Korean, and calculates word error for English and other Languages.

\subsection{Seed-ASR (CN)}
\label{sec:seed_asr_cn}
Seed-ASR (CN) follows the complete training pipeline shown in Figure \ref{train_stage}. In the SSL stage, we utilize the LUISE encoder with nearly 2B parameters, and conduct training on nearly eight million hours of Mandarin and Chinese dialect speech data from various domains. In the SFT stage, we use the trained LUISE and a MoE LLM with over ten billion parameters for model initialization. The training data comprises a mixture of Mandarin data containing multiple domain and dialect data. The detailed data distribution in the stage of SSL and SFT is introduced in Appendix \ref{stat_training_data}. In the Context SFT stage, we use a certain proportion of SFT-stage data mixed with some <context, speech, text> triple data for training. In the RL stage, we use the trained context SFT model for initialization, and construct high-quality training data for training. Following this comprehensive training process, we obtain Seed-ASR (CN).

To comprehensively evaluate the ASR ability of the Seed-ASR (CN) model, we compare it with other released models on public datasets and construct a series of evaluation sets, including the multi-domain sets, multi-source video sets, hardcase sets, multi-dialect sets, multi-accent sets, context-aware sets, and subjective intelligibility evaluation.

\subsubsection{Evaluation on Public Set}
We compare Seed-ASR (CN) with recently released large models on several Chinese ASR benchmarks, including: 1) the test set of aishell-1 \cite{bu2017aishell}, marked as aishell1\_test, with about 5 hours of read speech; 2) three test sets of aishell-2 \cite{du2018aishell}, marked as aishell2\_andriod, aishell2\_ios, and aishell2\_mic, each set containing about 5 hours of read speech; 3) two test sets of Wenetspeech \cite{zhang2022wenetspeech}, marked as wenetspeech\_testnet and wenetspeech\_testmeeting, containing 23 hours and 15 hours of multi-domain test data, respectively.

\begin{table}[h]
\caption{The comparison of Seed-ASR (CN) and other released large ASR models on Chinese ASR benchmarks. }
\label{public_set}
\centering
\resizebox{\textwidth}{!}{
\begin{tabular}{c|c|c|c|c}
\toprule
  &  Paraformer-large  &  Qwen-Audio  &  Hubert+Baichuan2  &  Seed-ASR (CN)  \\
\midrule
aishell1\_test  &  1.68  &  1.3  &  0.95  &  \textbf{0.68}   \\
aishell2\_andriod  &  3.13  &  3.3  &  \multirow{3}{*}{3.5 (avg)}  &  \textbf{2.27}   \\
aishell2\_ios  &  2.85  &  3.1  &  &  \textbf{2.27}  \\
aishell2\_mic  &  3.06  &  3.3  &  &  \textbf{2.28}  \\
wenetspeech\_testnet  &  6.74  &  9.5  &  6.06  &  \textbf{4.66}  \\
wenetspeech\_testmeeting  &  6.97  &  10.87  &  6.26  &  \textbf{5.69}  \\
\midrule
Average-6  &  4.07  &  5.23  &  3.96  &  \textbf{2.98}  \\
\bottomrule
\end{tabular}
}
\end{table}

The final result is the average of WER (character for Chinese) of the above 6 test sets. Our baselines for comparison include Paraformer-Large \cite{gao2022paraformer}, Qwen-Audio \cite{chu2023qwen}, and a recently released LLM-based ASR model with the structure of Hubert+Baichuan2 \cite{geng2024unveiling}. Their results presented here are from their respective papers. As shown in Table \ref{public_set}. Seed-ASR (CN) demonstrates a significant performance advantage over other models, achieving state-of-the-art results on these public datasets. For the average WER on the 6 sets, Seed-ASR (CN) achieves more than 24\%-40\% WER reduction than other published models.

\subsubsection{Evaluation on Multi-domain and Multi-source Video Set}
\label{cn_md}
We also conduct a comprehensive performance comparison on the multi-domain evaluation set, which contains high-quality evaluation data from various scenarios including video, live, voice search, meeting, intelligent assistants, etc. The weighted average WER of total 7 sets in multi-domain sets is used as the final metric. We select a transducer-based end-to-end models \cite{graves2012sequence} with a MoE encoder and over 300M parameters as one of the baselines. Additionally, we also run the results of Paraformer-large (offline decoding) on the multi-domain evaluation set as another baseline. From the results in Table \ref{basic_model_capability}, Seed-ASR (CN) shows significant performance advantage, with a relative decrease of more than 47\% in the WER metric compared to our strong end-to-end model. On the video evaluation sets covering 7 different subsets, Seed-ASR (CN) also obtains considerable performance improvement. These results demonstrate the strong foundational capabilities of Seed-ASR (CN).

\begin{table}[h]
\caption{Evaluation results on three sets covering multi-domain, multi-source video and hardcase with proper nouns. The metric of WER is used for the first two sets, and the F1 score of given keyword is used as the metric of hardcase set.}
\label{basic_model_capability}
\centering
\begin{tabular}{l|c|c|c}
\toprule
Model   &  \makecell[c]{Multidomain \\ (WER\%) $\downarrow$} &  \makecell[c]{Video-avg7 \\ (WER\%) $\downarrow$} & \makecell[c]{Hardcase \\ (F1\%) $\uparrow$} \\
\midrule
Transducer-based E2E Model  &  3.68  &  3.92  & 90.42  \\
Paraformer-large  &  5.23  &  5.97  & 87.99  \\
Seed-ASR (CN)  &  \textbf{1.94}  &  \textbf{2.70}  &  \textbf{93.72} \\
\bottomrule
\end{tabular}
\end{table}

Additionally, we evaluate the high-level ASR capabilities by introducing 10 hardcase test sets that cover utterances contain book titles, car names, idioms, drug names, movie names, ancient poems, product names, music names, etc. These test sets are designed to evaluate the model's ability to recognize speech content containing proper nouns with strong professionalism and domain specificity, reflecting the ASR model’s knowledge reserve and recognition accuracy. The evaluation metric for the hardcase sets is the F1 score of the given keyword in each sentence. As shown in Table \ref{basic_model_capability}, the Seed-ASR (CN) model achieves a 3.3\% absolute increase in the F1 value compared to the end-to-end model baseline, demonstrating the effectiveness of the AcLLM model framework in leveraging LLM's common sense knowledge and semantic reasoning capability.

\subsubsection{Evaluation on Multi-dialect Set and Multi-accent Set}
Since our Seed-ASR (CN) model supports the recognition of Mandarin and 13 Chinese dialects, we also introduce a dialect evaluation set. This set includes a total of 13 dialects (Cantonese, Southwest, Wu, Ji-lu, Zhongyuan, Min, etc.) and uses the same or similar pronunciation of Chinese characters to manually label the text. Specific demos of our dialect evaluation set are available on our website\footnote{\url{https://bytedancespeech.github.io/seedasr\_tech\_report}\label{website}}. We use WER as the objective metric for this dialect evaluation set.

\begin{table}[h]
\caption{Comparison on the 13 Chinese dialect evaluation sets.}
\label{multi_dialect}
\centering
\begin{tabular}{l|c}
\toprule
Model  &  \makecell[c]{Average WER \\ on 13 Chinese Dialects}  \\
\midrule
Finetuned Whisper Medium-v2  &  21.68   \\
Seed-ASR (CN)  &  19.09   \\
\bottomrule
\end{tabular}
\end{table}

We utilize a fine-tuned Whisper Medium-v2 with 769M parameters as our baseline. For a fair comparison, we train both Whisper Medium-v2 and Seed-ASR (CN) with the same dialect training set. Seed-ASR (CN) needs to maintain comprehensive capabilities in Mandarin while improving ASR performance on dialects, thus it is trained with a larger proportion of Mandarin data from multiple domains. In contrast, Whisper Medium-v2 shows inferior results on comprehensive evaluation sets such as the multi-domain set. Despite this, the Seed-ASR (CN) model, with its larger modeling capacity, still shows performance advantages over the baseline on the 13 dialect sets, with the average WER across the 13 dialects decreasing from 21.68 to 19.2 (an 11.4\% relative WER reduction), and a relative WER reduction of more than 21\% on a single dialect test set.

\begin{table}[h]
\caption{Comparison on the 11 Chinese accent evaluation sets.}
\label{multi_accent}
\centering
\begin{tabular}{l|c}
\toprule
Model  &  \makecell[c]{Average WER \\ on 11 Chinese Accents}  \\
\midrule
Transducer-based E2E Model  &  13.74   \\
Seed-ASR (CN) (w/o accent SFT data)  &  5.90   \\
Seed-ASR (CN)  &  4.96   \\
\bottomrule
\end{tabular}
\end{table}

To further verify the recognition performance of Seed-ASR (CN) on diverse speech, we introduce a series of accent evaluation sets, which include 11 Chinese accents from Anhui, Fujian, Gansu, Guangdong, Guizhou, Hunan, Jiangxi, Liaoning, Shaanxi, Shanxi, and Yunnan. Specific accent speech samples are also available on our website\textsuperscript{\ref{website}}. As shown in Table \ref{multi_accent}, Seed-ASR (CN) exhibits significant improvement on the accent test sets compared to our strong E2E model trained from scratch. We also conduct an ablation study by removing the accent SFT data during the training process, yet Seed-ASR (CN) still achieves strong performance on the accent sets. The results on multi-dialect and multi-accent evaluation sets demonstrate the strong robustness of Seed-ASR (CN) in recognizing Chinese speech from different regions.

\subsubsection{Evaluation on Dialogue Context Set}
\begin{figure}[h] 
\centering 
\includegraphics[width=\textwidth]{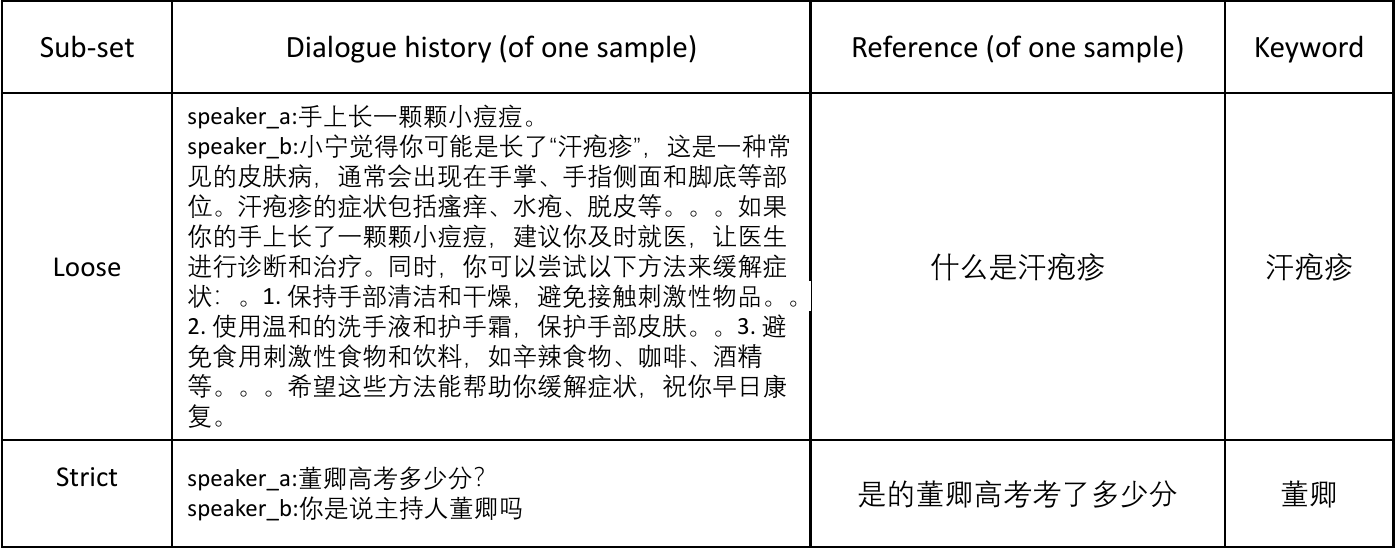}
\caption{Examples of strict and loose evaluation subsets.}
\label{strict_loose_intro} 
\end{figure}

In the evaluation of context awareness, we construct a high-quality dialogue context set where dialogue history is used as the contextual information. As shown in Figure \ref{strict_loose_intro}, we provide two examples of dialogues. Each test case includes the corresponding dialogue history text and the current recognized speech content. We divide the dialogue context evaluation into two subsets: strict and loose. The strict subset contains samples that have a strong dependence on the historical dialogue to accurately recognize the content of speech, such as person names. The loose subset has a weaker dependence between the historical dialogue and the content of speech, such as proper nouns. We use keyword recall as the evaluation metric.

On the dialogue evaluation set,  Seed-ASR (CN) model shows better keyword recall than a strong end-to-end transducer-based model that utilizes context FST biasing \cite{he2019streaming, zhao2019shallow} to improve keyword recall. Compared with Seed-ASR (CN) model that infers without context, the usage of context information brings more than a 15\% recall improvement. These results demonstrate the strong ability of our AcLLM model framework in utilizing the context-awareness capabilities of LLM.

\begin{table}[h]
\caption{The comparison of Seed-ASR and end-to-end models on our dialogue context sets, which cover strict subset and loose subset. Different decoding strategies are also compared.}
\label{rl_ablation_study}
\centering
\begin{tabular}{l|l|c}
\toprule
Model  &  Decoding method & \makecell[c]{Dialogue Context Set \\ Strict | Loose}  \\
\midrule
Transducer-based E2E Model  &  Context FST biasing  &  72.77 | 84.58   \\
Seed-ASR (CN)  &  Beam Search (w/o contexts)  &  65.45 | 89.33   \\
Seed-ASR (CN)  &  Joint Beam Search  &  \textbf{80.63} | \textbf{93.89}   \\
\bottomrule
\end{tabular}
\end{table}

On our website\textsuperscript{\ref{website}}, we also provide several demos showcasing the context-aware capabilities of Seed-ASR (CN). In the application scenario of intelligent assistants, the contexts not only include conversation history but also support information such as bot names, bot descriptions, and subtitle history. Additionally, we found that contextual information such as user edit history for video captions and the names of participants in meetings can also enhance the performance of Seed-ASR in their respective applications. 

\subsubsection{Subjective Evaluation}
In addition to the objective evaluations mentioned above, we also conduct a subjective evaluation to further measure the effectiveness of the Seed-ASR (CN) model. We selecte three well-educated transcribers to transcribe the audio in 5 test scenarios in the multidomain set (videos, live, voice search, meetings, and intelligent assistants). During transcription, the transcribers could listen to the samples multiple times and use search engines to ensure the accuracy of their transcription. After they complete the transcription, we will randomize the results from both the transcribers and the Seed-ASR (CN) model for subjective evaluation. The subjective evaluation metric is intelligibility, and the covered score range is 1-5 points. The scoring standard is shown in the following Figure \ref{subjective_standard}.
\begin{figure}[] 
\centering 
\includegraphics[width=\textwidth]{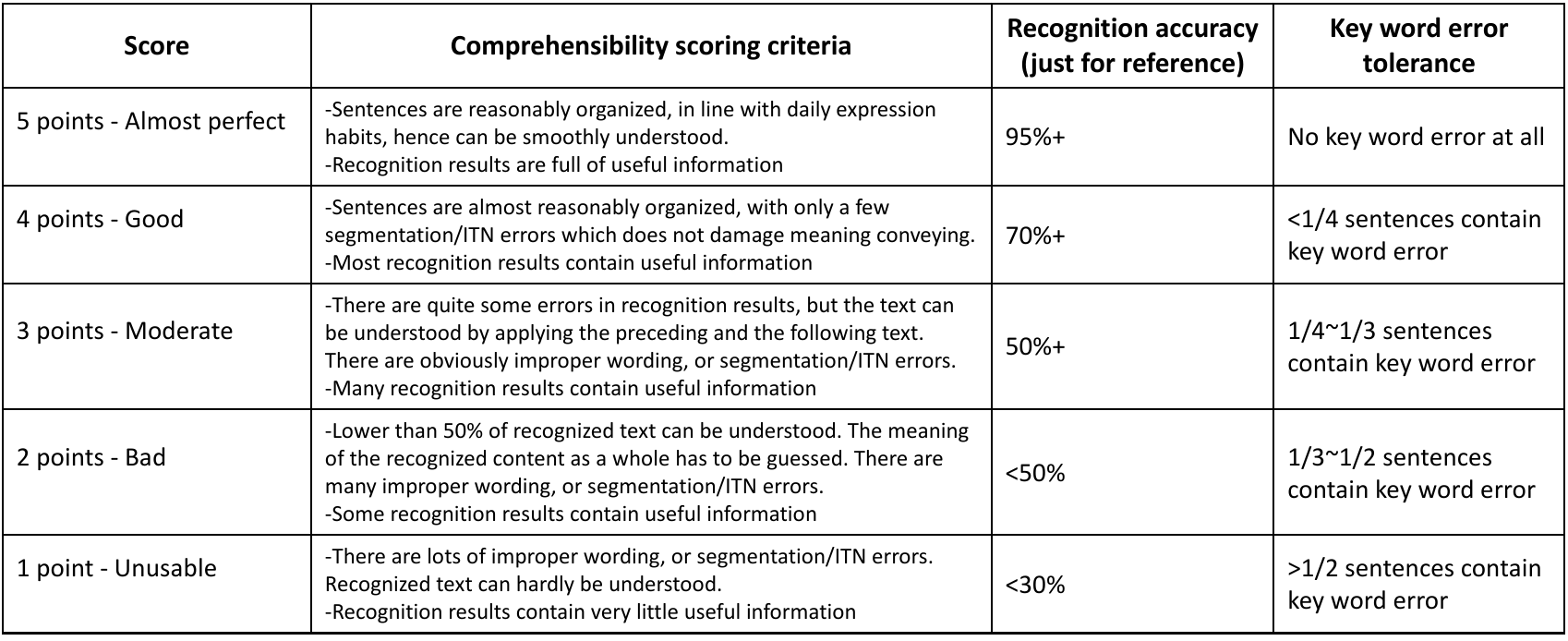}
\caption{The scoring standard for subjective evaluation.}
\label{subjective_standard} 
\end{figure}

On the test sets for voice search and voice assistants, the intelligibility of human recognition results is comparable to that of the Seed-ASR (CN) model. However, in live, videos, and meetings, Seed-ASR (CN) demonstrates better subjective intelligibility than humans. Specifically, compared to humans, in the case of professional field vocabulary and complex audio environments, the model can transcribe the content more accurately and give recognition results with higher intelligibility  compared with human.
\begin{table}[h]
\caption{Comparison of subjective intelligibility score between Seed-ASR (CN) and three human transcribers.}
\label{overall_ablation_study}
\centering
\resizebox{\textwidth}{!}{
\begin{tabular}{l|c|c|c|c|c|c|c}
\toprule
&  Voice search & Live & Video & Meeting & Intelligent assistant & Average \\
\midrule
3 Human Results  &  4.89/4.85/4.87  &  4.26/4.58/4.50  &  4.60/4.64/4.63  &  4.30/4.03/4.37  &  4.92/4.85/4.88  &  -  \\
Human Average &  4.87  &  4.45  &  4.62  &  4.23  &  4.88  &  4.61  \\
Seed-ASR (CN) &  4.9  &  4.81  &  4.89  &  4.76  &  4.92  &  4.86  \\
\bottomrule
\end{tabular}
}
\end{table}

\subsubsection{Summary}
Following a stage-by-stage training recipe including SFT $\rightarrow$ context SFT $\rightarrow$ RL, our Seed-ASR (CN) model is produced. On above comprehensive evaluation sets, we observe that certain capabilities of our Seed-ASR (CN) model are enhanced at different training stages. Here, we present a detailed ablation study on the effect of each stage, with results shown in Table \ref{overall_ablation_study}. First, the introduction of the RL stage brings improvements on most evaluation sets, such as multi-domain, multi-source video, multi-dialect, hardcase, and code-switch. The slight degradation in the accent test set may be due to the training data ratio. Additionally, training in the context SFT stage positively impacts most test sets, notably bringing significant improvement in the recall metric on the context strict test set. This further demonstrates the effectiveness of our context-aware training and decoding strategy in the context SFT stage.

\begin{table}[h]
\caption{Ablation studies on Seed-ASR (CN) after different stages.}
\label{overall_ablation_study}
\centering
\resizebox{\textwidth}{!}{
\begin{tabular}{l|c|c|c|c|c|c|c}
\toprule
Model & \makecell[c]{Multi-domain \\ (WER\%) $\downarrow$} & \makecell[c]{Multi-source \\ Video \\ (WER\%) $\downarrow$} & \makecell[c]{Multi-accent \\ (WER\%) $\downarrow$} & \makecell[c]{Multi-dialect \\ (WER\%) $\downarrow$} & \makecell[c]{Hardcase \\ (F1\%) $\uparrow$} & \makecell[c]{Context-strict \\ (recall\%) $\uparrow$} & \makecell[c]{Code-switch \\ (WER\%) $\downarrow$} \\
\midrule
Seed-ASR (CN)  &  \textbf{1.94}  &  \textbf{2.7}  &  4.96  &  \textbf{19.09}  &  \textbf{93.72}  &  \textbf{80.63}  &  \textbf{5.65}  \\
\quad w/o RL  &  2.02  &  2.79  &  5.05  &  19.48  &  93.39  &  80.63  &  6.07  \\
\quad\quad w/o context SFT &  2.11  &  2.82  &  \textbf{4.89}  &  19.47  &  93.43  &  61.26  &  5.93  \\
\bottomrule
\end{tabular}
}
\end{table}

The evaluation results demonstrate that Seed-ASR (CN) possesses more comprehensive and powerful model capabilities compared to classic end-to-end models and other released models. The performance advantage of Seed-ASR is evident in public test sets and our subjective intelligibility evaluation, where it even surpasses human transcribers in some domains. Moreover, Seed-ASR has achieved significant recall improvements compared to end-to-end models combined with context FST fusion strategies on the context-aware evaluation set. This unified and concise structure reflects Seed-ASR’s ability to support customized ASR application scenarios. Overall, the evaluation results showcase the powerful capabilities of the Seed-ASR model in various ASR scenarios that handle diverse speech inputs and contexts.

\subsection{Seed-ASR (ML)}
As demonstrated above, Seed-ASR (CN) exhibits strong performance in recognizing Mandarin and Chinese dialects. To extend these advantages to languages spoken by users in other countries, we also apply the Seed-ASR methodology to multilingual scenarios, resulting in our multilingual model: Seed-ASR (ML). The training of Seed-ASR (ML) differs from Seed-ASR (CN) primarily in terms of the training data. While Seed-ASR (CN) focuses on Mandarin and Chinese dialects, Seed-ASR (ML) is trained on a diverse set of multilingual data. In the stage of SSL, the audio encoder of Seed-ASR (ML) also utilizes the LUISE with 2B parameters, and is trained with over tens of millions of hours of unsupervised multilingual data from multi-domain sources. In the subsequent stages, we select the training data from our multilingual ASR training sets sum up to hundreds of thousands of hours covering 9 languages: English, Chinese, Arabic, Spanish, French, Indonesian, Japanese, Korean and Portuguese. The detailed data distribution in the stage of SSL and SFT is introduced in Appendix \ref{stat_training_data}. We conduct performance comparisons on our multiple evaluation sets and public datasets.

\subsubsection{Evaluation on Multi-domain and Multi-accent Sets}
On the multi-domain evaluation sets, the covered domains are the same as the multi-domain evaluation sets on Seed-ASR (CN) introduced in Section \ref{cn_md}. The hardcase test sets cover domains ranging from medical health, food and drink, sports, technology, outfit, games, entertainment and beauty. We also build an evaluation of different accents of English, including speakers from Great Britain, United States, Australia, Canada, China, India, Singapore, New Zealand and South Africa. For multilingual evaluation, we report the average WER performance on 7 non-English languages: Arabic (AR), Spanish (ES), French (FR), Indonesian (ID), Japanese (JA), Korean (KO), and Portuguese (PT). As shown in Table \ref{ml_comparison}, the baselines for comparison include Google USM\cite{zhang2023google}  (API call \footnote{https://sites.research.google/usm/}), Whisper Large v3\cite{radford2023robust} (offline decoding) and Universal-1\cite{ramirez2024anatomy} (API call \footnote{https://www.assemblyai.com/app/}). Since Universal-1 only supports 3 languages in our multilingual multi-domain evaluation sets, its corresponding results are not included here. We attach the language-wise performance comparison on multilingual multi-domain evaluation sets among these models to Appendix \ref{appendix_ml}. From the results in Table \ref{ml_comparison}, Seed-ASR (ML) demonstrates relatively over 42\% and 40\% on English and multilingual multi-domain evaluation sets, respectively, compared to the strongest baselines. Similar significant improvements are also observed on the English multi-accent and hardcase evaluation sets.

\begin{table}[h]
\caption{Comparison with Google USM, Whisper Large v3 and Universal-1 on English multi-domain, multi-accent, hardcase evaluation sets, and multilingual multi-domain evaluation sets.}
\label{ml_comparison}
\centering
\resizebox{\textwidth}{!}{
\begin{tabular}{c|c|c|c|c|c}
\toprule
&  Google USM\cite{zhang2023google}  &  Whisper Large v3\cite{radford2023robust} & Universal-1\cite{ramirez2024anatomy}  & Seed-ASR (ML)  \\
\midrule
English Multi-domain (WER\%) $\downarrow$  &  9.33  &  10.41 &  9.95  &  \textbf{5.34}  \\
\hline
English Multi-accent (WER\%) $\downarrow$ & 22.19 & 21.52 & 14.40 & \textbf{11.26} \\
\hline
English Hardcase (F1\%) $\uparrow$ & 63.30 & 79.54 & 77.82 & \textbf{87.94} \\
\hline \hline
Multilingual Multi-domain (WER\%) $\downarrow$  &  21.51  &  20.55 & -  &  \textbf{12.16}   \\
\bottomrule
\end{tabular}
}
\end{table}

\subsubsection{Evaluation on Public Sets}
In addition to the internal multi-domain evaluation sets, we also compare Seed-ASR (ML) with other models on public test sets for English and other languages, including Librispeech \cite{panayotov2015librispeech} test clean/other, MLS \cite{pratap2020mls}, Tedlium 3 \cite{hernandez2018ted}, Callhome, Switchboard\cite{godfrey1992switchboard}, AMI \cite{kraaij2005ami}, and Fleurs \cite{conneau2023fleurs}. Details of the test sets are introduced in Appendix \ref{appendix_ml_Public}. The results are illustrated in Table \ref{ml_public}. Note that all the results of baseline models are WERs reported by the respective papers or technical reports of the baseline models (Whisper Large-v3 results are from the Universal-1's technical report \cite{ramirez2024anatomy}).  As shown in Table \ref{ml_public}, Seed-ASR (ML) achieves top performance on most of the test sets across different languages with improvements ranging from 10\% to 40\%, indicating Seed-ASR (ML)'s generalization ability to domains unseen during training.

\begin{table}[h]
\caption{ASR Results of Seed-ASR (ML) on English and Multilingual Public test sets}
\label{ml_public}
\centering
\resizebox{\textwidth}{!}{
\begin{tabular}{c|c|c|c|c|c|c|c}
\toprule
Test set  &  Language & Google USM\cite{zhang2023google}  &  Whisper Large-v2\cite{radford2023robust} &  Whisper Large-v3  &  Universal-1\cite{ramirez2024anatomy} & Gemini-1.5 Pro\cite{reid2024gemini}  & Seed-ASR (ML)  \\
\midrule
Librispeech test\_clean & EN &  -  &  2.7  &  1.8  &  1.6  &  -  &  \textbf{1.58}  \\
Librispeech test\_other & EN &  -  &  5.2  &  3.6  &  3.1  &  -  & \textbf{2.84}   \\
Tedlium 3 & EN &  -  &  4.0  &  7.4  &  7.5  &  -  &  \textbf{3.11}  \\
Switchboard & EN &  -  &  13.8  &  -  &  -  &  -  &  \textbf{11.59}  \\
CallHome & EN &  -  &  17.6  &  -  &  -  &  -  &  \textbf{12.24}  \\
AMI IHM & EN &  -  &  16.9  &  -   &  -  &  -  &  \textbf{13.16}  \\
\midrule
\multirow{7.0}{*}{Fleurs} &  EN  & - &  4.4  &  -  &  - & -  &  \textbf{3.43}  \\
&  AR & -  &  16  &  -  &  -  & - &  \textbf{13.05}  \\
&  ES & - &  3.0  &  2.8  &  5.0  & - & \textbf{2.50}  \\
&  FR & - &  8.3  &  \textbf{5.6}  &  6.8  & - & 7.09  \\
&  ID & - &  7.1  &  -  &  -  & - & \textbf{4.24}  \\
&  JA & - &  5.3  &  -  &  -  & - & \textbf{3.46}  \\
&  KO & - &  14.3  &  -  &  -  & - & \textbf{3.25}  \\
&  PT & - &  4.3  &  -  &  -  & - & \textbf{3.55}  \\
\midrule
\multirow{3}{*}{MLS} &  EN & 7 &  6.2  &  -  &  - & 4.6 &  \textbf{4.14}  \\
&  ES & - &  4.2  &  5.7  &  \textbf{3.3} & - &  3.76  \\
&  FR & - &  7.3  &  8.1  &  \textbf{2.3} & - &  5.10  \\
&  PT & - &  6.8  &  -  &  -  & - & \textbf{5.04}  \\
\midrule
\bottomrule
\end{tabular}
}
\end{table}

\subsubsection{Summary}
Similar with Seed-ASR (CN), Seed-ASR (ML) demonstrates exceptional performance across a wide range of evaluation sets comapred to multiple strong baselines. The model excels in recognizing speech with diverse acoustic environments, semantic contexts and accents across multiple languages, underscoring the model’s generalization ability and its effectiveness in processing speech from various unseen domains during training. Overall, the results on above evaluation sets on Chinese and multilingual setting demonstrate the generalization and strong foundation abilities of Seed-ASR in diverse application scenarios covering multi-lingual, multi-dialect, multi-accent, multi-domain, and multiple customization requirements.

\section{Conclusion}
The Seed-ASR models, trained through a stage-by-stage recipe including SFT, context SFT, and RL, demonstrates superior capabilities across various evaluation sets across different acoustic and semantic domains, accents/dialects/languages, and long range speech duration, compared to recently-released strong end-to-end models. The large-scale LUISE pretraining and SFT to connect LUISE and LLM endow Seed-ASR capacity to understand diverse speech content. The introduction of context SFT stage significantly boosts the models' recall on keywords given related context, showcasing the model's strong customization ability in utilizing the context-awareness abilities of LLMs. The RL stage further consolidates the alignment between Seed-ASR's text generation behavior and the requirement for accurate transcription, especially the transcription of semantically important parts. Overall, the results affirm Seed-ASR’s position as a top-performing ASR model for diverse applications involving multiple languages, dialects, accents, domains, and customization needs. In future, we will focus on extending Seed-ASR's ability to deal with multiple tasks within a single model, further enhance the long-form ability and increase the number of supported languages.

\newpage


\bibliographystyle{plain}
\bibliography{Seed-ASR}

\begin{thebibliography}{10}

\bibitem{anil2023palm}
Rohan Anil, Andrew~M Dai, Orhan Firat, Melvin Johnson, Dmitry Lepikhin, Alexandre Passos, Siamak Shakeri, Emanuel Taropa, Paige Bailey, Zhifeng Chen, et~al.
\newblock Palm 2 technical report.
\newblock {\em arXiv preprint arXiv:2305.10403}, 2023.

\bibitem{baevski2020wav2vec}
Alexei Baevski, Yuhao Zhou, Abdelrahman Mohamed, and Michael Auli.
\newblock wav2vec 2.0: A framework for self-supervised learning of speech representations.
\newblock {\em Advances in neural information processing systems}, 33:12449--12460, 2020.

\bibitem{bahdanau2016end}
Dzmitry Bahdanau, Jan Chorowski, Dmitriy Serdyuk, Philemon Brakel, and Yoshua Bengio.
\newblock End-to-end attention-based large vocabulary speech recognition.
\newblock In {\em 2016 IEEE international conference on acoustics, speech and signal processing (ICASSP)}, pages 4945--4949. IEEE, 2016.

\bibitem{brown2020language}
Tom Brown, Benjamin Mann, Nick Ryder, Melanie Subbiah, Jared~D Kaplan, Prafulla Dhariwal, Arvind Neelakantan, Pranav Shyam, Girish Sastry, Amanda Askell, et~al.
\newblock Language models are few-shot learners.
\newblock {\em Advances in neural information processing systems}, 33:1877--1901, 2020.

\bibitem{bu2017aishell}
Hui Bu, Jiayu Du, Xingyu Na, Bengu Wu, and Hao Zheng.
\newblock Aishell-1: An open-source mandarin speech corpus and a speech recognition baseline.
\newblock In {\em 2017 20th conference of the oriental chapter of the international coordinating committee on speech databases and speech I/O systems and assessment (O-COCOSDA)}, pages 1--5. IEEE, 2017.

\bibitem{chan2016listen}
William Chan, Navdeep Jaitly, Quoc Le, and Oriol Vinyals.
\newblock Listen, attend and spell: A neural network for large vocabulary conversational speech recognition.
\newblock In {\em 2016 IEEE international conference on acoustics, speech and signal processing (ICASSP)}, pages 4960--4964. IEEE, 2016.

\bibitem{chen2023x}
Feilong Chen, Minglun Han, Haozhi Zhao, Qingyang Zhang, Jing Shi, Shuang Xu, and Bo~Xu.
\newblock X-llm: Bootstrapping advanced large language models by treating multi-modalities as foreign languages.
\newblock {\em arXiv preprint arXiv:2305.04160}, 2023.

\bibitem{chen2022wavlm}
Sanyuan Chen, Chengyi Wang, Zhengyang Chen, Yu~Wu, Shujie Liu, Zhuo Chen, Jinyu Li, Naoyuki Kanda, Takuya Yoshioka, Xiong Xiao, et~al.
\newblock Wavlm: Large-scale self-supervised pre-training for full stack speech processing.
\newblock {\em IEEE Journal of Selected Topics in Signal Processing}, 16(6):1505--1518, 2022.

\bibitem{chen2024salm}
Zhehuai Chen, He~Huang, Andrei Andrusenko, Oleksii Hrinchuk, Krishna~C Puvvada, Jason Li, Subhankar Ghosh, Jagadeesh Balam, and Boris Ginsburg.
\newblock Salm: Speech-augmented language model with in-context learning for speech recognition and translation.
\newblock In {\em ICASSP 2024-2024 IEEE International Conference on Acoustics, Speech and Signal Processing (ICASSP)}, pages 13521--13525. IEEE, 2024.

\bibitem{chiu2022self}
Chung-Cheng Chiu, James Qin, Yu~Zhang, Jiahui Yu, and Yonghui Wu.
\newblock Self-supervised learning with random-projection quantizer for speech recognition.
\newblock In {\em International Conference on Machine Learning}, pages 3915--3924. PMLR, 2022.

\bibitem{chowdhery2023palm}
Aakanksha Chowdhery, Sharan Narang, Jacob Devlin, Maarten Bosma, Gaurav Mishra, Adam Roberts, Paul Barham, Hyung~Won Chung, Charles Sutton, Sebastian Gehrmann, et~al.
\newblock Palm: Scaling language modeling with pathways.
\newblock {\em Journal of Machine Learning Research}, 24(240):1--113, 2023.

\bibitem{chu2023qwen}
Yunfei Chu, Jin Xu, Xiaohuan Zhou, Qian Yang, Shiliang Zhang, Zhijie Yan, Chang Zhou, and Jingren Zhou.
\newblock Qwen-audio: Advancing universal audio understanding via unified large-scale audio-language models.
\newblock {\em arXiv preprint arXiv:2311.07919}, 2023.

\bibitem{conneau2023fleurs}
Alexis Conneau, Min Ma, Simran Khanuja, Yu~Zhang, Vera Axelrod, Siddharth Dalmia, Jason Riesa, Clara Rivera, and Ankur Bapna.
\newblock Fleurs: Few-shot learning evaluation of universal representations of speech.
\newblock In {\em 2022 IEEE Spoken Language Technology Workshop (SLT)}, pages 798--805. IEEE, 2023.

\bibitem{devlin2018bert}
Jacob Devlin, Ming-Wei Chang, Kenton Lee, and Kristina Toutanova.
\newblock Bert: Pre-training of deep bidirectional transformers for language understanding.
\newblock {\em arXiv preprint arXiv:1810.04805}, 2018.

\bibitem{dong2020cif}
Linhao Dong and Bo~Xu.
\newblock Cif: Continuous integrate-and-fire for end-to-end speech recognition.
\newblock In {\em ICASSP 2020-2020 IEEE International Conference on Acoustics, Speech and Signal Processing (ICASSP)}, pages 6079--6083. IEEE, 2020.

\bibitem{du2018aishell}
Jiayu Du, Xingyu Na, Xuechen Liu, and Hui Bu.
\newblock Aishell-2: Transforming mandarin asr research into industrial scale.
\newblock {\em arXiv preprint arXiv:1808.10583}, 2018.

\bibitem{gao2022paraformer}
Zhifu Gao, Shiliang Zhang, Ian McLoughlin, and Zhijie Yan.
\newblock Paraformer: Fast and accurate parallel transformer for non-autoregressive end-to-end speech recognition.
\newblock {\em arXiv preprint arXiv:2206.08317}, 2022.

\bibitem{geng2024unveiling}
Xuelong Geng, Tianyi Xu, Kun Wei, Bingsheng Mu, Hongfei Xue, He~Wang, Yangze Li, Pengcheng Guo, Yuhang Dai, Longhao Li, et~al.
\newblock Unveiling the potential of llm-based asr on chinese open-source datasets.
\newblock {\em arXiv preprint arXiv:2405.02132}, 2024.

\bibitem{godfrey1992switchboard}
John~J Godfrey, Edward~C Holliman, and Jane McDaniel.
\newblock Switchboard: Telephone speech corpus for research and development.
\newblock In {\em Acoustics, speech, and signal processing, ieee international conference on}, volume~1, pages 517--520. IEEE Computer Society, 1992.

\bibitem{graves2012sequence}
Alex Graves.
\newblock Sequence transduction with recurrent neural networks.
\newblock {\em arXiv preprint arXiv:1211.3711}, 2012.

\bibitem{graves2006connectionist}
Alex Graves, Santiago Fern{\'a}ndez, Faustino Gomez, and J{\"u}rgen Schmidhuber.
\newblock Connectionist temporal classification: labelling unsegmented sequence data with recurrent neural networks.
\newblock In {\em Proceedings of the 23rd international conference on Machine learning}, pages 369--376, 2006.

\bibitem{gulati2020conformer}
Anmol Gulati, James Qin, Chung-Cheng Chiu, Niki Parmar, Yu~Zhang, Jiahui Yu, Wei Han, Shibo Wang, Zhengdong Zhang, Yonghui Wu, et~al.
\newblock Conformer: Convolution-augmented transformer for speech recognition.
\newblock {\em arXiv preprint arXiv:2005.08100}, 2020.

\bibitem{he2019streaming}
Yanzhang He, Tara~N Sainath, Rohit Prabhavalkar, Ian McGraw, Raziel Alvarez, Ding Zhao, David Rybach, Anjuli Kannan, Yonghui Wu, Ruoming Pang, et~al.
\newblock Streaming end-to-end speech recognition for mobile devices.
\newblock In {\em ICASSP 2019-2019 IEEE International Conference on Acoustics, Speech and Signal Processing (ICASSP)}, pages 6381--6385. IEEE, 2019.

\bibitem{hernandez2018ted}
Fran{\c{c}}ois Hernandez, Vincent Nguyen, Sahar Ghannay, Natalia Tomashenko, and Yannick Esteve.
\newblock Ted-lium 3: Twice as much data and corpus repartition for experiments on speaker adaptation.
\newblock In {\em Speech and Computer: 20th International Conference, SPECOM 2018, Leipzig, Germany, September 18--22, 2018, Proceedings 20}, pages 198--208. Springer, 2018.

\bibitem{hinton2012deep}
Geoffrey Hinton, Li~Deng, Dong Yu, George~E Dahl, Abdel-rahman Mohamed, Navdeep Jaitly, Andrew Senior, Vincent Vanhoucke, Patrick Nguyen, Tara~N Sainath, et~al.
\newblock Deep neural networks for acoustic modeling in speech recognition: The shared views of four research groups.
\newblock {\em IEEE Signal processing magazine}, 29(6):82--97, 2012.

\bibitem{hoffmann2022training}
Jordan Hoffmann, Sebastian Borgeaud, Arthur Mensch, Elena Buchatskaya, Trevor Cai, Eliza Rutherford, Diego de~Las Casas, Lisa~Anne Hendricks, Johannes Welbl, Aidan Clark, et~al.
\newblock Training compute-optimal large language models.
\newblock {\em arXiv preprint arXiv:2203.15556}, 2022.

\bibitem{hsu2021hubert}
W.~Hsu, B.~Bolte, Y.~H. Tsai, K.~Lakhotia, R.~Salakhutdinov, and A.~Mohamed.
\newblock {HuBERT}: Self-supervised speech representation learning by masked prediction of hidden units.
\newblock {\em IEEE/ACM Transactions on Audio, Speech, and Language Processing}, 29:3451--3460, 2021.

\bibitem{huang2024audiogpt}
Rongjie Huang, Mingze Li, Dongchao Yang, Jiatong Shi, Xuankai Chang, Zhenhui Ye, Yuning Wu, Zhiqing Hong, Jiawei Huang, Jinglin Liu, et~al.
\newblock Audiogpt: Understanding and generating speech, music, sound, and talking head.
\newblock In {\em Proceedings of the AAAI Conference on Artificial Intelligence}, volume~38, pages 23802--23804, 2024.

\bibitem{kaplan2020scaling}
Jared Kaplan, Sam McCandlish, Tom Henighan, Tom~B Brown, Benjamin Chess, Rewon Child, Scott Gray, Alec Radford, Jeffrey Wu, and Dario Amodei.
\newblock Scaling laws for neural language models.
\newblock {\em arXiv preprint arXiv:2001.08361}, 2020.

\bibitem{kraaij2005ami}
Wessel Kraaij, Thomas Hain, Mike Lincoln, and Wilfried Post.
\newblock The ami meeting corpus.
\newblock In {\em Proc. International Conference on Methods and Techniques in Behavioral Research}, 2005.

\bibitem{li2023prompting}
Y.~Li, Y.~Wu, J.~Li, and S.~Liu.
\newblock Prompting large language models for zero-shot domain adaptation in speech recognition.
\newblock {\em arXiv:2306.16007}, 2023.

\bibitem{miao2015eesen}
Yajie Miao, Mohammad Gowayyed, and Florian Metze.
\newblock Eesen: End-to-end speech recognition using deep rnn models and wfst-based decoding.
\newblock In {\em 2015 IEEE workshop on automatic speech recognition and understanding (ASRU)}, pages 167--174. IEEE, 2015.

\bibitem{cass2012language}
Chinese~Academy of~Social~Sciences and City~University of~Hong~Kong.
\newblock the language atlas of china.
\newblock {\em The Commercial Press}, 2012.

\bibitem{openai2022chatgpt}
OpenAI.
\newblock Introducing chatgpt.
\newblock {\em URL https://openai.com/blog/chatgpt}, 2022.

\bibitem{openai2023gpt4}
OpenAI.
\newblock Gpt-4 technical report.
\newblock 2023.

\bibitem{panayotov2015librispeech}
Vassil Panayotov, Guoguo Chen, Daniel Povey, and Sanjeev Khudanpur.
\newblock Librispeech: an asr corpus based on public domain audio books.
\newblock In {\em 2015 IEEE international conference on acoustics, speech and signal processing (ICASSP)}, pages 5206--5210. IEEE, 2015.

\bibitem{prabhavalkar2018minimum}
Rohit Prabhavalkar, Tara~N Sainath, Yonghui Wu, Patrick Nguyen, Zhifeng Chen, Chung-Cheng Chiu, and Anjuli Kannan.
\newblock Minimum word error rate training for attention-based sequence-to-sequence models.
\newblock In {\em 2018 IEEE International Conference on Acoustics, Speech and Signal Processing (ICASSP)}, pages 4839--4843. IEEE, 2018.

\bibitem{pratap2020mls}
Vineel Pratap, Qiantong Xu, Anuroop Sriram, Gabriel Synnaeve, and Ronan Collobert.
\newblock Mls: A large-scale multilingual dataset for speech research.
\newblock {\em arXiv preprint arXiv:2012.03411}, 2020.

\bibitem{radford2023robust}
Alec Radford, Jong~Wook Kim, Tao Xu, Greg Brockman, Christine McLeavey, and Ilya Sutskever.
\newblock Robust speech recognition via large-scale weak supervision.
\newblock In {\em International Conference on Machine Learning}, pages 28492--28518. PMLR, 2023.

\bibitem{radford2019language}
Alec Radford, Jeffrey Wu, Rewon Child, David Luan, Dario Amodei, Ilya Sutskever, et~al.
\newblock Language models are unsupervised multitask learners.
\newblock {\em OpenAI blog}, 1(8):9, 2019.

\bibitem{ramirez2024anatomy}
Francis~McCann Ramirez, Luka Chkhetiani, Andrew Ehrenberg, Robert McHardy, Rami Botros, Yash Khare, Andrea Vanzo, Taufiquzzaman Peyash, Gabriel Oexle, Michael Liang, et~al.
\newblock Anatomy of industrial scale multilingual asr.
\newblock {\em arXiv preprint arXiv:2404.09841}, 2024.

\bibitem{reid2024gemini}
Machel Reid, Nikolay Savinov, Denis Teplyashin, Dmitry Lepikhin, Timothy Lillicrap, Jean-baptiste Alayrac, Radu Soricut, Angeliki Lazaridou, Orhan Firat, Julian Schrittwieser, et~al.
\newblock Gemini 1.5: Unlocking multimodal understanding across millions of tokens of context.
\newblock {\em arXiv preprint arXiv:2403.05530}, 2024.

\bibitem{rubenstein2023audiopalm}
P.K. Rubenstein, C.~Asawaroengchai, D.D. Nguyen, et~al.
\newblock {AudioPaLM}: {A} large language model that can speak and listen.
\newblock {\em arXiv:2306.12925}, 2023.

\bibitem{tang2023salmonn}
Changli Tang, Wenyi Yu, Guangzhi Sun, Xianzhao Chen, Tian Tan, Wei Li, Lu~Lu, Zejun Ma, and Chao Zhang.
\newblock Salmonn: Towards generic hearing abilities for large language models.
\newblock {\em arXiv preprint arXiv:2310.13289}, 2023.

\bibitem{touvron2023llama}
Hugo Touvron, Thibaut Lavril, Gautier Izacard, Xavier Martinet, Marie-Anne Lachaux, Timoth{\'e}e Lacroix, Baptiste Rozi{\`e}re, Naman Goyal, Eric Hambro, Faisal Azhar, et~al.
\newblock Llama: Open and efficient foundation language models.
\newblock {\em arXiv preprint arXiv:2302.13971}, 2023.

\bibitem{touvron2023llama2}
Hugo Touvron, Louis Martin, Kevin Stone, Peter Albert, Amjad Almahairi, Yasmine Babaei, Nikolay Bashlykov, Soumya Batra, Prajjwal Bhargava, Shruti Bhosale, et~al.
\newblock Llama 2: Open foundation and fine-tuned chat models.
\newblock {\em arXiv preprint arXiv:2307.09288}, 2023.

\bibitem{watanabe2017hybrid}
Shinji Watanabe, Takaaki Hori, Suyoun Kim, John~R Hershey, and Tomoki Hayashi.
\newblock Hybrid ctc/attention architecture for end-to-end speech recognition.
\newblock {\em IEEE Journal of Selected Topics in Signal Processing}, 11(8):1240--1253, 2017.

\bibitem{wu2023decoder}
J.~Wu, Y.~Gaur, Z.~Chen, et~al.
\newblock On decoder-only architecture for speech-to-text and large language model integration.
\newblock {\em arXiv:2307.03917}, 2023.

\bibitem{zhang2022wenetspeech}
Binbin Zhang, Hang Lv, Pengcheng Guo, Qijie Shao, Chao Yang, Lei Xie, Xin Xu, Hui Bu, Xiaoyu Chen, Chenchen Zeng, et~al.
\newblock Wenetspeech: A 10000+ hours multi-domain mandarin corpus for speech recognition.
\newblock In {\em ICASSP 2022-2022 IEEE International Conference on Acoustics, Speech and Signal Processing (ICASSP)}, pages 6182--6186. IEEE, 2022.

\bibitem{zhang2023google}
Yu~Zhang, Wei Han, James Qin, Yongqiang Wang, Ankur Bapna, Zhehuai Chen, Nanxin Chen, Bo~Li, Vera Axelrod, Gary Wang, et~al.
\newblock Google usm: Scaling automatic speech recognition beyond 100 languages.
\newblock {\em arXiv preprint arXiv:2303.01037}, 2023.

\bibitem{zhao2019shallow}
Ding Zhao, Tara~N Sainath, David Rybach, Pat Rondon, Deepti Bhatia, Bo~Li, and Ruoming Pang.
\newblock Shallow-fusion end-to-end contextual biasing.
\newblock In {\em Interspeech}, pages 1418--1422, 2019.

\end{thebibliography}
\newpage

\section{Authors (alphabetical order)}
\vspace{-0.2cm}
\begin{table}[h]
\centering
\begin{tabular}{p{4cm}p{4cm}p{4cm}}
Ye Bai&Mingkun Huang&Ming Tu\\
Jingping Chen&Youjia Huang&Bo Wang\\
Jitong Chen&Jishuo Jin&Hao Wang\\
Wei Chen&Fanliu Kong&Yuping Wang\\
Zhuo Chen&Zongwei Lan&Yuxuan Wang\\
Chuang Ding&Tianyu Li&Hanzhang Xia\\
Linhao Dong&Xiaoyang Li&Rui Xia\\
Qianqian Dong&Zeyang Li&Shuangyi Xie\\
Yujiao Du&Zehua Lin&Hongmin Xu\\
Kepan Gao&Rui Liu&Meng Yang\\
Lu Gao&Shouda Liu&Bihong Zhang\\
Yi Guo&Lu Lu&Jun Zhang\\
Minglun Han&Yizhou Lu&Wanyi Zhang\\
Ting Han&Jingting Ma&Yang Zhang\\
Wenchao Hu&Shengtao Ma&Yawei Zhang\\
Xinying Hu&Yulin Pei&Yijie Zheng\\
Yuxiang Hu&Chen Shen&Ming Zou\\
Deyu Hua&Tian Tan&\\
Lu Huang&Xiaogang Tian&\\
\end{tabular}
\end{table}
\vspace{-0.2cm}

\appendix

\section{Appendix}

\subsection{Detailed Results of Seed-ASR (ML)}
\label{appendix_ml}
In Table \ref{ml_lang_res}, we present a language-wise comparison among Google USM, Whisper Large-v3, and Seed-ASR (ML) on the multilingual multi-domain evaluation sets for non-English languages. The results clearly demonstrate Seed-ASR (ML)’s advantage in every language, with WER reduction ranging from 26\% to 47\%. For the two relatively low-resource languages, Arabic (AR) and Indonesian (ID), which are spoken by large populations in the world, Seed-ASR (ML) achieves a relative WER reduction of over 45\%.

\begin{table}[h]
\caption{Language-wise performance of Seed-ASR (ML) on multilingual multi-domain evaluation sets.}
\label{ml_lang_res}
\centering
\resizebox{0.6\textwidth}{!}{
\begin{tabular}{c|c|c|c}
\toprule
Language & Google USM  &  Whisper Large-v3  & Seed-ASR (ML)  \\
\midrule
AR &   35.21  & 48.31 &  \textbf{18.69}  \\
ES &   15.20  & 16.68 & \textbf{10.28}  \\
FR & 20.48  & 17.62 & \textbf{12.70}  \\
ID &   22.29  & 20.47 & \textbf{10.86}  \\
JA &   24.62  & 18.57 & \textbf{13.72}  \\
KO &  13.88  & 13.07 & \textbf{7.77}  \\
PT &  19.88  & 18.78 & \textbf{11.69}  \\
\midrule
\bottomrule
\end{tabular}
}
\end{table}

\subsection{Details of English and Multilingual public test sets used in Seed-ASR (ML) evaluation}
\label{appendix_ml_Public}
The details of the English and multilingual public test sets are as follows:

\textbf{Librispeech} \cite{panayotov2015librispeech}: as per usual, we report the WER on the test-clean and test-other sets.

\textbf{Tedlium 3} \cite{hernandez2018ted}: The test set provided in Tedlium 3 with segmented transcripts is employed.

\textbf{CallHome, Switchboard and AMI IHM}: we keep consistent with Whisper v3 \cite{radford2023robust} by using the two corpora from LDC2002S09 and LDC2002T43 for CallHome and Switchboard. We only report the IHM subset from AMI corpus.

\textbf{Fleurs} \cite{conneau2023fleurs}: since transcripts from Fleurs test sets are annotated and processed with text normalization, we asked linguistics to conduct the inverse text normalization for the 8 languages and calculated the WERs on the corresponding transcripts. 

\textbf{MLS} \cite{pratap2020mls}: we evaluated the test subset of English, Spanish, French and Portuguese from MLS.

\subsection{Training Dataset Statistics}
\label{stat_training_data}

In this section, we present the statistical information regarding the amount and language of data used in self-supervised learning (SSL) of LUISE and supervised fine-tuning (SFT) of Seed-ASR. This includes four parts: the speech-only data used in the training of LUISE used in Seed-ASR (CN) and Seed-ASR (ML), and the general ASR data used for Seed-ASR (CN) and Seed-ASR (ML).

\begin{figure}[h] 
\centering 
\includegraphics[width=0.7\textwidth, trim=10cm 3cm 4cm 1cm, clip]{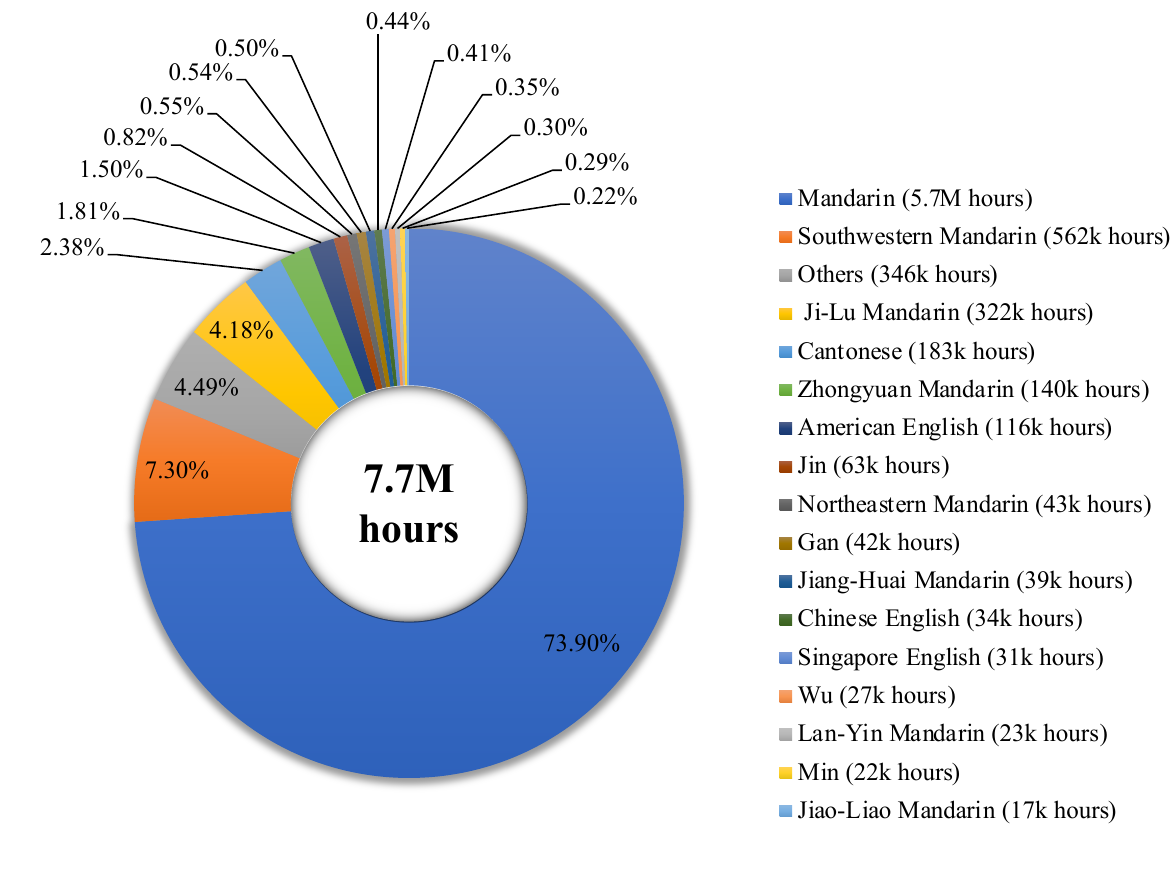}
\setlength{\belowcaptionskip}{-10pt}
\caption{Training data statistics of the large-scale self-supervised learning of LUISE used in Seed-ASR (CN). (1) The total amount of training data is 7.7 million hours; (2) Mandarin Chinese has the highest proportion with about 5.6 million hours of speech data, accounting for about 74\%; In addition to Mandarin Chinese, we also include other Chinese dialects and categorize them according to the Language Atlas of China \cite{cass2012language}; (3) We also include English data from different regions, as well as a small amount of Mandarin-English code-switching data.}
\label{context_exp} 
\end{figure}

\begin{figure}[h] 
\centering 
\includegraphics[width=0.7\textwidth, trim=10cm 4cm 6cm 2cm, clip]{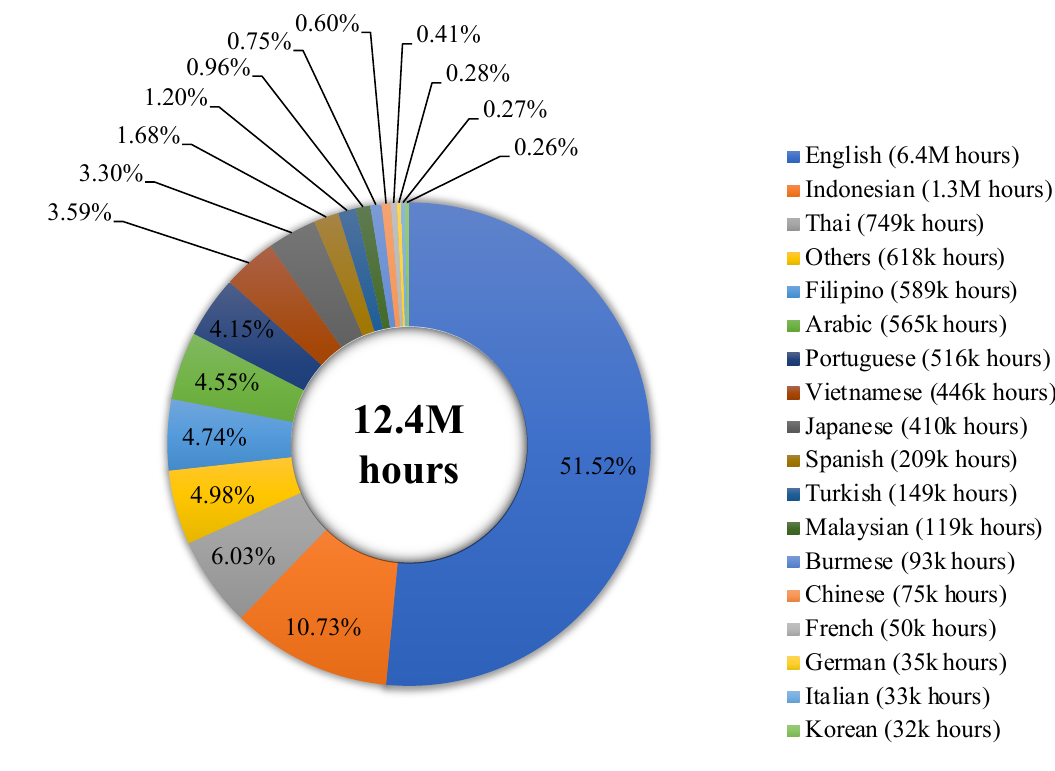}
\setlength{\belowcaptionskip}{-10pt}
\caption{Training data statistics of the large-scale self-supervised learning of LUISE used in Seed-ASR (ML). (1) The total amount of training data is 12.4 million hours; (2) English has the highest proportion with about 6.4 million hours speech data, accounting for 51.52\%; (3) In addition to English, we also include data from more than 20 other languages; (4) Categories with less than 0.2\% of the data (such as Romanian, Polish, Dutch, Russian, etc.) were grouped into the "Others".}
\label{context_exp} 
\end{figure}

\begin{figure}[h] 
\centering 
\includegraphics[width=0.8\textwidth, trim=11cm 4cm 3cm 2cm, clip]{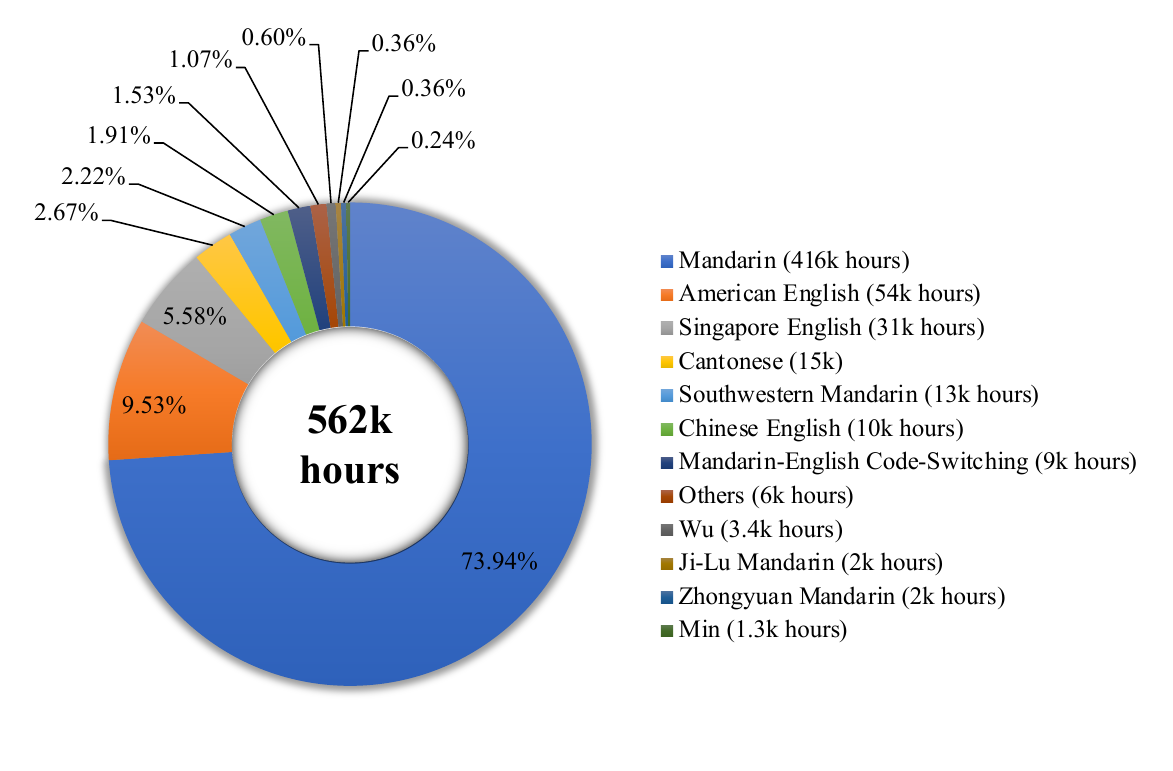}
\caption{Training data statistics of the supervised fine-tuning of Seed-ASR (CN). (1) The total amount of training data is 562k hours; (2) Mandarin Chinese has the highest proportion with about 416k hours speech data, accounting for 73.94\%; (3) In addition to Mandarin Chinese, we also include data of Chinese dialects and English from different regions; (4) Categories with less than 0.2\% of the data (such as Jin, Min, Xiang, Hakka, etc.) were grouped into the "Others".}
\label{context_exp} 
\end{figure}

\begin{figure}[h] 
\centering 
\includegraphics[width=0.8\textwidth, trim=10cm 4cm 4cm 2cm, clip]{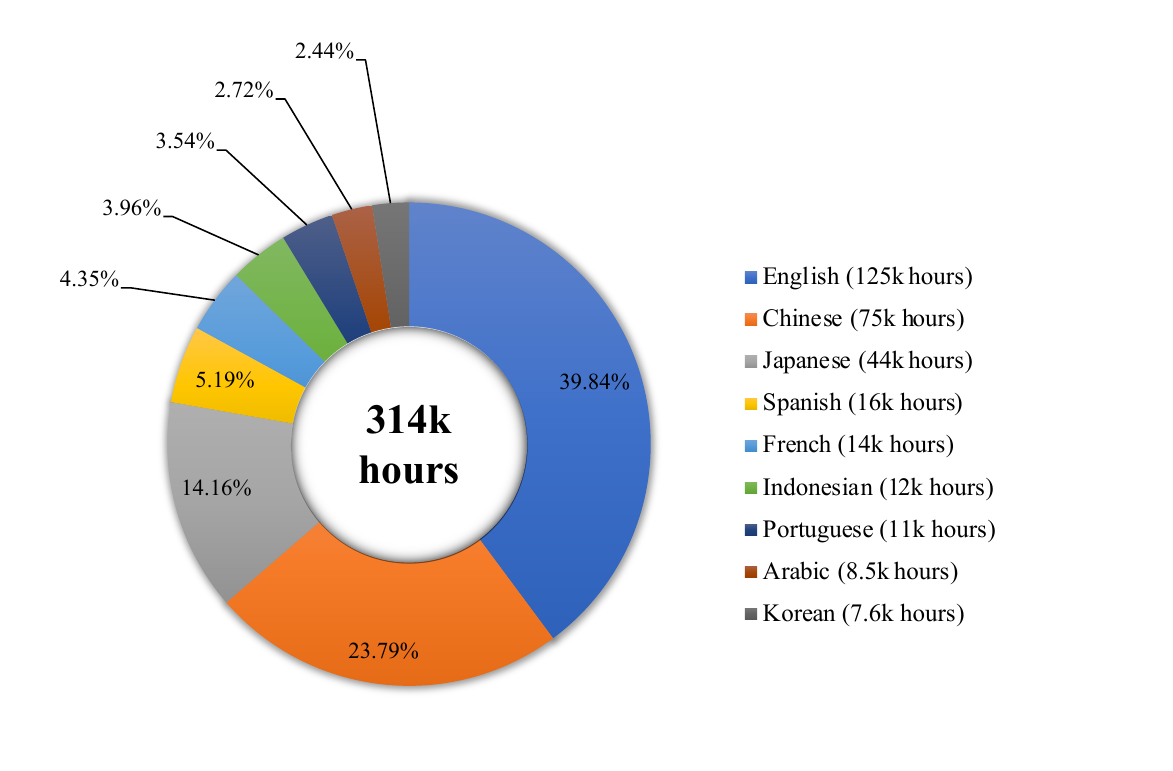}
\caption{Training data statistics of the supervised fine-tuning of Seed-ASR (ML). (1) The total amount of training data is 314k hours; (2) English has the highest proportion with about 125k hours speech data, accounting for 39.84\%; (3) In addition to English, we include some languages that are widely used around the world, as mentioned in the main text.}
\label{context_exp} 
\end{figure}

\end{spacing}

\end{document}